\documentclass[%
reprint,
amsmath,amssymb,
aps,
]{revtex4-2}

\usepackage{graphicx}
\usepackage{dcolumn}
\usepackage{bm}
\usepackage[colorlinks=true,citecolor=black,urlcolor=black,linkcolor=black,filecolor=black]{hyperref}

\usepackage[capitalise]{cleveref}
\graphicspath{{figs/}}
\usepackage{mathtools}

\usepackage[defaultcolor=red]{changes}
\usepackage[multiple]{footmisc}

\begin{document}

\title{Slow Voltage Relaxation of Silicon Nanoparticles with a Chemo-Mechanical Core-Shell Model}


\author{Lukas Köbbing}
\author{Yannick Kuhn}
\author{Birger Horstmann}
\email{birger.horstmann@dlr.de}
\affiliation{Institute of Engineering Thermodynamics, German Aerospace Center (DLR), Wilhelm-Runge-Straße 10, 89081 Ulm, Germany}
\affiliation
{Helmholtz Institute Ulm (HIU), Helmholtzstraße 11, 89081 Ulm, Germany}
\affiliation{Faculty of Natural Sciences, Ulm University, Albert-Einstein-Allee 47, 89081 Ulm, Germany}

\begin{abstract}

Silicon presents itself as a high-capacity anode material for lithium-ion batteries with a promising future. The high ability for lithiation comes along with massive volume changes and a problematic voltage hysteresis, causing reduced efficiency, detrimental heat generation, and a complicated state-of-charge estimation. During slow cycling, amorphous silicon nanoparticles show a larger voltage hysteresis than after relaxation periods. Interestingly, the voltage relaxes for at least several days, which has not been physically explained so far. We apply a chemo-mechanical continuum model in a core-shell geometry interpreted as a silicon particle covered by the solid-electrolyte interphase to account for the hysteresis phenomena. The silicon core (de)lithiates during every cycle while the covering shell is chemically inactive. The visco-elastoplastic behavior of the shell explains the voltage hysteresis during cycling and after relaxation. We identify a logarithmic voltage relaxation, which fits with the established Garofalo law for viscosity. Our chemo-mechanical model describes the observed voltage hysteresis phenomena and outperforms the empirical Plett model. In addition to our full model, we present a reduced model to allow for easy voltage profile estimations. The presented results support the mechanical explanation of the silicon voltage hysteresis with a core-shell model and encourage further efforts into the investigation of the silicon anode mechanics.

\vspace{1.2cm}

\end{abstract}

\maketitle

\section{Introduction}

For the enhancement of next-generation lithium-ion batteries, research and industry consider the application of pure silicon anodes \cite{Sun2022, Zuo2017, Feng2018}. Silicon is a popular choice as it is an abundant and cheap material. Anodes made from silicon possess a high theoretical capacity, leading to a massive volume expansion of up to $300\,\%$ during lithiation and respective shrinkage during delithiation \cite{Beaulieu2001}. The massive deformations induce significant stresses inside the anode material, causing fracture of large silicon particles above a critical diameter of $150\,\mathrm{nm}$ \cite{Liu2012}. Larger silicon particles suffer from cracks, particle pulverization, and are prone to losing contact with the current collector \cite{Wetjen2018}. Anodes made from silicon nanoparticles promise a higher stability and cycle life compared to anodes with larger silicon particles. Thus, research and industry focus on the behavior of nano-structured silicon anodes \cite{Kilchert2024}.

A severe challenge for the commercialization of silicon anodes is the handling and possible reduction of the voltage hysteresis observed in various experiments \cite{Pan2019, Pan2020, Bernard2019, Wycisk2023}. Silicon anodes reveal a different voltage during slow lithiation compared to delithiation reducing efficiency and causing detrimental heat generation \cite{McDowell2013, Wycisk2023Heat}. Experiments observe this hysteresis phenomenon of amorphous silicon anodes in thin-film geometries, micron-sized particles, and nanoparticles. Literature discusses different reasons for the voltage hysteresis: mechanics and plastic flow of silicon in thin-film geometries \cite{Sethuraman2010a, Lu2016}, concentration gradients due to slow diffusion in micrometer-sized particles \cite{Chandrasekaran2010, Zhao2011, Cui2012}, phase transformation in the very first cycle \cite{McDowell2013Trafo}, slow reaction kinetics \cite{Sethuraman2013}, and mechanical impact of the solid-electrolyte interphase on the lithiation behavior of silicon nanoparticles \cite{Koebbing2023Voltage}.

The solid-electrolyte interphase (SEI) forms on the surface of the silicon anode particles resulting from the electrochemical instability of the electrolyte in contact with the anode \cite{Horstmann2019, Zhang2021interplay, Nie2013}. This process leads to a significant capacity loss, even more pronounced for silicon nanoparticle anodes due to the high surface-to-volume ratio. The SEI film is only several nanometers thick but can effectively limit further electrolyte decomposition \cite{Verma2010, Peled2017}. Nevertheless, a leakage of electrons through the SEI causes a continued, long-term growth of the SEI during storage and cycling \cite{Koebbing2023Growth, Kolzenberg2020}. Concerning the massive volume changes of silicon particles during cycling, it is important to consider the mechanics of the SEI, namely elastic and plastic deformations as well as fracture \cite{Kolzenberg2021}. However, experimental and theoretical results indicate that the innermost layer of the SEI withstands the large stresses due to plastic flow and fast self-healing \cite{Guo2020, Lee2024, Kolzenberg2021}.
In addition to the SEI, a natural silicon oxide layer usually covers the surface of silicon particles and contributes to the innermost SEI layer \cite{Cao2019}.

For silicon nanoparticles, the SEI has a beneficial effect on their mechanical integrity \cite{Li2019, Chen2020, Sun2022}. Furthermore, carbon-coated silicon particles show a reduced voltage hysteresis \cite{Bernard2019}. Both observations support our recent explanation of the voltage hysteresis for silicon nanoparticles based on SEI mechanics \cite{Koebbing2023Voltage}.
Due to the reported pulverization of large particles \cite{Wetjen2018} and growth of SEI into the interior of silicon anodes \cite{He2021}, the mechanical explanation of the hysteresis for silicon nanoparticles covered by a surface layer also has the potential to reason the hysteresis observed for larger geometries \cite{Koebbing2023Voltage}. Nonetheless, recent experimental results reveal a slow voltage relaxation behavior for at least $300\,\mathrm{h}$ and a significantly enlarged cycling voltage hysteresis even at $\mathrm{C}/100$ not explained so far \cite{Wycisk2023}.
The slow voltage relaxation is in line with the experimental findings of Sethuraman et al. \cite{Sethuraman2013}. However, their theoretical explanation with reaction kinetics in the Tafel regime requires extreme values for the exchange current density and the transfer coefficients. Here, we present a consistent picture of the voltage relaxation and other voltage hysteresis phenomena based on mechanics.

This manuscript builds on our previous explanation of the voltage hysteresis of silicon nanoparticles by the mechanical interaction of silicon and SEI \cite{Koebbing2023Voltage}. We explain the basic principles of our chemo-mechanical core-shell model in \cref{sec:theory}. Furthermore, we introduce the Garofalo viscosity model necessary because of the large stresses arising inside the SEI shell and discuss its behavior in the core-shell system with an analytical approximation and a reduced model. In \cref{sec:experimental}, we describe the recent experiments performed by Wycisk et al. \cite{Wycisk2023}, which we analyze in detail in \cref{sec:results}. In conclusion, we present a consistent description of the observed slow voltage relaxation, hysteresis shape, C-rate dependence, and voltage transition profiles.

\section{Theory}
\label{sec:theory}

\begin{figure*}[htbp]
	\centering
	\includegraphics[width=1.0\textwidth]{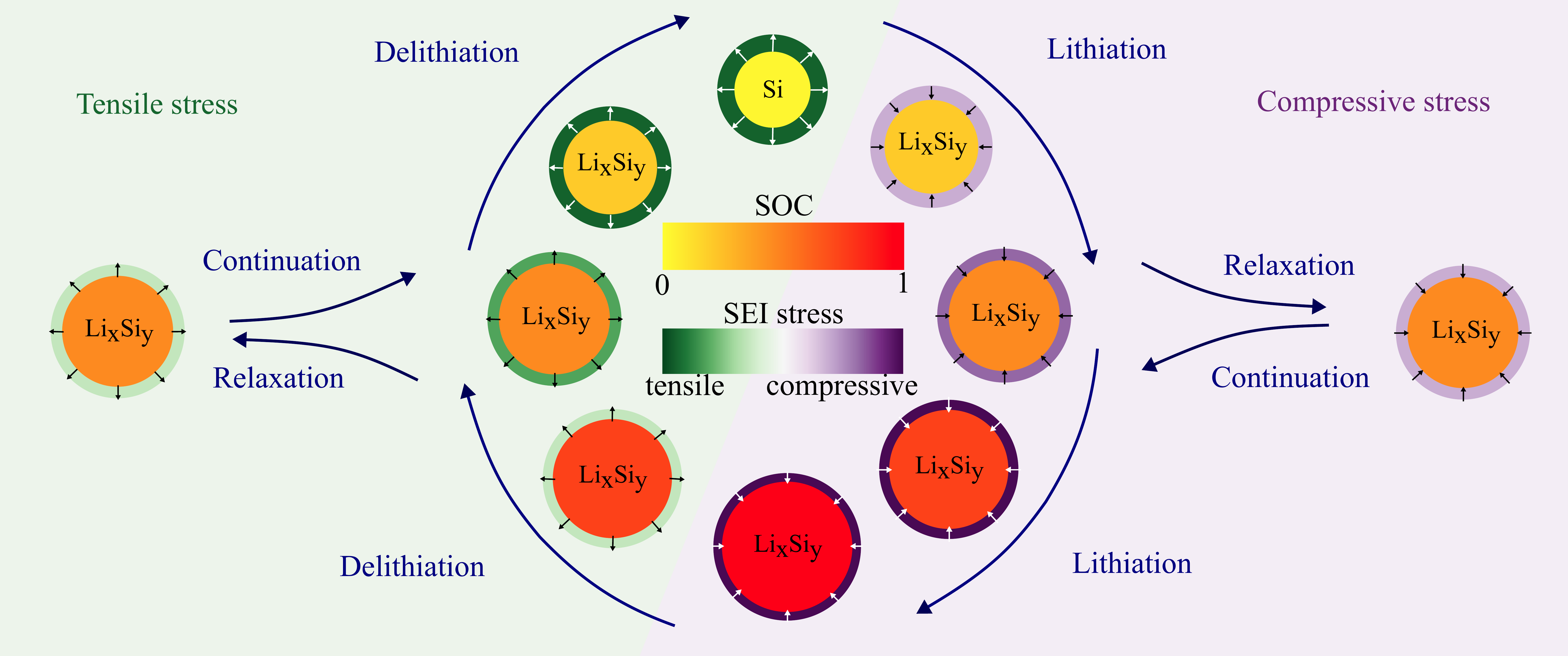}
	\caption{Scheme of volume change and SEI stress during lithiation, delithiation, and relaxation periods.}
	\label{fig:scheme}
\end{figure*}

Our theoretical framework describes the behavior of a core-shell system, where the silicon particle as core can lithiate and delithiate while the shell is chemically inactive and deforms only mechanically as illustrated in \cref{fig:scheme}. The core-shell model can in principle relate to different scenarios: (i) SEI on silicon nanoparticle or pulverized silicon micro-particle. Literature reports that the inner SEI on silicon is robust against the severe volume changes \cite{Guo2020, Lee2024, Kolzenberg2021}. (ii) Silicon oxide layer on silicon due to exposure to air or electrolyte, which is reported to affect the silicon lithiation behavior \cite{Schnabel2020, Schroder2014}. This effect potentially combines with the influence of the SEI \cite{Cao2019}. (iii) Silicon nanodomains in silicon particles surrounded by chemically inactive regions. Literature reports the existence of silicon nanodomains for amorphous silicon under high pressure \cite{Deringer2021}, for crystalline silicon \cite{Chevrier2014}, and generically for silicon oxide particles \cite{Kitada2019, Wang2020}. We interpret the core-shell behavior as a particle-SEI system in the following.

We have presented the foundations of the chemo-mechanical particle-SEI model used in this study in our previous publications \cite{Kolzenberg2021, Koebbing2023Voltage}. In the following, we summarize the most important assumptions and equations. Further, we highlight advancements compared to our previous works.

\subsection{Chemo-Mechanical Core-Shell Model}
\label{sec:model}

The silicon particle core deforms due to the chemical lithiation and delithiation $\mathbf{F}_\mathrm{core,ch}$, elastic deformation $\mathbf{F}_\mathrm{core,el}$, and plastic deformation $\mathbf{F}_\mathrm{core,pl}$ when reaching the yield criterion. The large deformation approach determines the total deformation $\mathbf{F}_\mathrm{core}$ as
\begin{equation}
	\mathbf{F}_\mathrm{core} = \mathbf{F}_\mathrm{core,pl} \mathbf{F}_\mathrm{core,el} \mathbf{F}_\mathrm{core,ch} .
	\label{eq:deformations}
\end{equation}
The concentration of lithium atoms $c_\mathrm{Li,0}$ inside the silicon particle expressed in the undeformed Lagrangian frame determines the chemical deformation \begin{equation}
	\mathbf{F}_\mathrm{core,ch} = \lambda_\mathrm{ch} \mathbf{Id} = (1+v_\mathrm{Li} c_\mathrm{Li,0})^{1/3} \mathbf{Id}
\end{equation}
with $v_\mathrm{Li}$ the molar volume of lithium inside silicon.

The strain tensors $\mathbf{E}_\mathrm{core,k}$ read
\begin{equation}
	\mathbf{E}_\mathrm{core,k} = \frac{1}{2} \left(\mathbf{F}_\mathrm{core,k}^\mathrm{T}\mathbf{F}_\mathrm{core,k}^{\phantom{\mathrm{T}}}-\mathbf{Id}\right),
	\label{eq:strains}
\end{equation}
where the subscript $\mathrm{k}$ indicates the kind of deformation, which is either the total deformation or one of the mentioned deformation contributions from \cref{eq:deformations}.

The Cauchy stress $\sigma$ describes the stress in the deformed Euler frame and the Piola-Kirchhoff stress $\mathbf{P} = J \sigma \mathbf{F}^{-\mathrm{T}}$ describes the stress in the undeformed Lagrangian frame with $J=\operatorname{det} \mathbf{F}$. The Piola-Kirchhoff stress due to elastoplastic deformation reads
\begin{equation}
\begin{aligned}
    \mathbf{P}_\mathrm{core} = \lambda_\mathrm{ch}^{-2} &\mathbf{F}^\mathrm{\phantom{+1}}_\mathrm{core} \mathbf{F}_\mathrm{core,pl}^{-\mathrm{T}} \mathbf{F}_\mathrm{core,pl}^{-1} \\
    &\left(\lambda_\mathrm{core} \operatorname{tr}(\mathbf{E}_\mathrm{core,el})\mathbf{Id} + 2G_\mathrm{core} \mathbf{E}_\mathrm{core,el}\right)
\end{aligned}
 \label{eq:stress-elastoplastic}
\end{equation}
with the first Lamé constant $\lambda_\mathrm{core}$ and the second Lamé constant $G_\mathrm{core}$.

Due to the chemo-mechanical coupling, the stress inside the particle affects the voltage as
\begin{equation}
	U = U_0 + \frac{v_\mathrm{Li}}{3 F \lambda_\mathrm{ch}^3} \mathbf{P}_\mathrm{core} : \mathbf{F}_\mathrm{core}
	\label{eq:ocv}
\end{equation}
with the true open-circuit voltage (OCV) of silicon $U_0$ and the Faraday constant $F$.

The differential equations of interest inside the particle are the continuity equation for the time derivative of the lithium concentration $\dot{c}_{\mathrm{Li}, 0}$, the momentum balance, and the equation for the plastic flow rate $\dot{\mathbf{F}}_\mathrm{core,pl}$,
\begin{align}
    \dot{c}_{\mathrm{Li}, 0} &=-\nabla_{0} \cdot \vec{N}_{\mathrm{Li}, 0}\label{eq:dot-concentration}\\
    0 &= \nabla_{0} \cdot \mathbf{P}_\mathrm{core}\label{eq:nabla-stress}\\
    \dot{\mathbf{F}}^\mathrm{\phantom{+1}}_\mathrm{core,pl} \mathbf{F}_\mathrm{core,pl}^{-1} &= \phi_\mathrm{core} \frac{\partial f_\mathrm{core}}{\partial \mathbf{M}_\mathrm{core}}.\label{eq:plastic-flow}
\end{align}
For the lithiation equation, we define the lithium flux $\vec{N}_{\mathrm{Li}, 0} = -L \nabla_0 \mu_\mathrm{Li}$, the electro-chemo-mechanical potential $\mu_\mathrm{Li}=-FU$, the mobility $L=D_\mathrm{Li} \left(\partial \mu_\mathrm{Li}/\partial c_\mathrm{Li,0}\right)^{-1}$, and the diffusion coefficient $D_\mathrm{Li}$. At the particle boundary, the (de)lithiation rate determines the lithium flux $\vec{N}_{\mathrm{Li}, 0}(R)$. For the plastic flow, the von Mises yield criterion $f_\mathrm{core} = \frac{3}{2} |\mathbf{M}^\mathrm{dev}_\mathrm{core}|^2 / \sigma_\mathrm{Y,core}^2 - 1 \leq 0$ determines plasticity with $\mathbf{M}^\mathrm{dev}_\mathrm{core} = \mathbf{M}_\mathrm{core} - 1/3\operatorname{tr} \mathbf{M}_\mathrm{core}$ the deviatoric part of the adapted Mandel stress $\mathbf{M}_\mathrm{core} = \mathbf{F}_\mathrm{core,rev}^\mathrm{T} \sigma_\mathrm{core} \mathbf{F}_\mathrm{core,rev}^{-\mathrm{T}}$, the reversible deformation $\mathbf{F}_\mathrm{core,rev}=\mathbf{F}_\mathrm{core,el}\mathbf{F}_\mathrm{core,ch}$, and the yield stress $\sigma_\mathrm{Y,core}$. The consistency condition $\dot{f}_\mathrm{core} = 0$ determines the plastic multiplier $\phi_\mathrm{core}$.

For the shell behavior, we assume that the SEI shell deforms only mechanically, namely elastically and plastically,
\begin{equation}
	\mathbf{F}_\mathrm{shell} = \mathbf{F}_\mathrm{shell,pl} \mathbf{F}_\mathrm{shell,el}
\end{equation}
leading to massive mechanical strains and stresses when experiencing the significant volume change of the silicon particle during cycling. Analogous to the particle, \cref{eq:strains} determines the strain tensors $\mathbf{E}_{\mathrm{shell,k}}$ inside the SEI.

In addition to the elastoplastic stress $\mathbf{P}_\mathrm{shell,el}$ determined analogously to \cref{eq:stress-elastoplastic}, we consider the viscous behavior of the SEI. To describe large viscous stresses during cycling on the one hand and small viscous stresses during relaxation on the other hand, we use the Garofalo law or inverse hyperbolic sine law
\begin{equation}
	\mathbf{P}_{\mathrm{shell,visc}} = J_\mathrm{shell} \sigma_\mathrm{ref} \cdot \operatorname{asinh}\left(\tau \dot{\mathbf{E}}_{\mathrm{shell}}\right) \mathbf{F}_\mathrm{shell}^{-\mathrm{T}}
	\label{eq:Garofalo-law}
\end{equation}
calculated component-wise and presented initially in Ref. \cite{Garofalo1963}. The parameter $\sigma_\mathrm{ref}$ describes as a reference stress the magnitude of viscous stress at a certain strain rate. The parameter $\tau$ describes the time constant of the system and the dependence on the strain rate. In this study, we use the Garofalo viscosity model stated in \cref{eq:Garofalo-law} instead of a standard Newtonian model or a shear-thinning model \cite{Koebbing2023Voltage} to account more adequately for the complexity of the mechanical behavior. The particular functional dependence of the Garofalo law is reasoned in Ref. \cite{Stang2018} by a change in the energy landscape due to mechanical deformations and lattice distortions. Furthermore, positive entropy production is guaranteed analogously to the derivation in Ref. \cite{Koebbing2023Voltage} as the inverse hyperbolic sine is positive for positive arguments and negative for negative ones.

The differential equations of interest inside the SEI shell are the momentum balance and the equation for plastic flow,
\begin{align}
0 &= \nabla_{0} \cdot \left( \mathbf{P}_\mathrm{shell,el} + \mathbf{P}_\mathrm{shell,visc} \right)
 \label{eq:nabla-stress-sei}\\
 \dot{\mathbf{F}}^\mathrm{\phantom{+1}}_\mathrm{shell,pl} \mathbf{F}_\mathrm{shell,pl}^{-1} &= \phi_\mathrm{shell} \frac{\partial f_\mathrm{shell}}{\partial \mathbf{M}_\mathrm{shell,el}}.
    \label{eq:plastic-flow-sei}
\end{align}
The yield criterion $f_\mathrm{shell} = \frac{3}{2} |\mathbf{M}_\mathrm{shell,el}^\mathrm{dev}|^2 / \sigma_\mathrm{Y,shell}^2 - 1 \leq 0$ is determined by the deviatoric part $\mathbf{M}_\mathrm{shell,el}^\mathrm{dev}$ of the adapted elastic Mandel stress $\mathbf{M}_\mathrm{shell,el} = \mathbf{F}_\mathrm{shell,el}^\mathrm{T} \sigma_\mathrm{shell,el} \mathbf{F}_\mathrm{shell,el}^{-\mathrm{T}}$ and the plastic multiplier $\phi_\mathrm{shell}$ results from the consistency condition $\dot{f}_\mathrm{shell}=0$.

Note that we model the mechanical deformations on a continuum scale. Thus, the visco-elastoplastic behavior is not necessarily an intrinsic property of a single material domain. Instead, interfaces and grain boundaries of multiple crystal domains can determine the continuum mechanics. Hence, the described visco-elastoplasticity can be a consequence of repeated partial cracking and healing, as discussed for the SEI in Ref. \cite{Kolzenberg2021}. This description is reasonable because the literature does not observe significant fracture of the inner SEI layer on silicon \cite{Guo2020, Lee2024, Kolzenberg2021}.

We assume that the surfaces of the silicon particle and SEI stick tightly together, meaning that the radial part of the stress coincides
\begin{equation}
	\mathbf{P}_\mathrm{core,rr} \big\rvert_{r=R} = \mathbf{P}_\mathrm{shell,rr} \big\rvert_{r=R}
	\label{eq:boundary-condition}
\end{equation}
when evaluated at the particle-SEI interface $r=R$. Due to the merely mechanical deformation of the SEI, significant stresses arise inside the SEI impacting the silicon particle stress and voltage.

As presented in Ref. \cite{Koebbing2023Voltage}, the expansion of the silicon particle during lithiation leads to a mechanical reaction of the SEI, namely, first elastic and then plastic deformation. The strains inside the SEI generate significant compressive stress acting on the silicon particle as visualized in \cref{fig:scheme}. Additionally, viscous behavior increases the total compressive stress during lithiation depending on the strain rate. During the subsequent delithiation, tensile stress originates from elastic and plastic deformations as well as viscosity, which causes a stress hysteresis inside the SEI, impacting the voltage of the silicon particle according to \cref{eq:ocv}. Hence, the visco-elastoplastic behavior of the SEI describes the voltage hysteresis observed for silicon nanoparticles.

\subsection{Analytical Approximation for the Voltage Relaxation}
\label{sec:analytical-solution}

To gain an analytical approximation for the voltage relaxation, we investigate the behavior of the presented chemo-mechanical core-shell model in a simplified setup. Thus, we analyze all local variables at the interface accounting for the central role of the interface coupling. In the following, we discuss several assumptions paving the way to a simplified analytical expression.

First, we choose the simplified description that during relaxation the silicon particle behaves purely elastically according to Hooke's law
\begin{equation}
	\sigma_\mathrm{ev} = E_\mathrm{core} \cdot \mathbf{E}_\mathrm{core,ev,rr}
	\label{eq:hooke}
\end{equation}
with Young's modulus $E_\mathrm{core}$ and the elastic strain of the core $\mathbf{E}_\mathrm{core,ev,rr}$ due to viscous stress of the shell.

Furthermore, we consider only the viscous stress contribution inside the shell as the elastic stress of the shell stays constant, i.e.,
\begin{equation}
	\sigma_\mathrm{shell,visc} = \sigma_\mathrm{ref} \cdot \mathrm{asinh}\left(\tau\dot{\mathbf{E}}_\mathrm{shell}\right).
	\label{eq:pde-visc-stress}
\end{equation}
The time evolution of the radial stress component in the silicon particle resulting from the time derivative of \cref{eq:hooke} states
\begin{equation}
	\frac{\mathrm{d}\sigma_\mathrm{ev}}{\mathrm{d}t} = E_\mathrm{core} \cdot \dot{\mathbf{E}}_\mathrm{core,ev,rr}.
	\label{eq:pde-time-derivative}
\end{equation}
The silicon core deforms only elastically during relaxation of viscous shell stress and isotropically, thus
\begin{equation}
	\dot{\mathbf{E}}_\mathrm{core,ev} \approx \dot{\mathbf{F}}_\mathrm{core,ev} \approx \frac{\dot{\mathbf{F}}_\mathrm{core}}{\lambda_\mathrm{ch}} \approx \frac{\dot{\mathbf{E}}_\mathrm{core}}{\lambda_\mathrm{ch}^2}
\end{equation}
and
\begin{equation}
	\dot{\mathbf{E}}_\mathrm{core,ev,rr} = \dot{\mathbf{E}}_\mathrm{core,ev,\varphi\varphi} = \frac{\dot{\mathbf{E}}_\mathrm{core,\varphi\varphi}}{\lambda_\mathrm{ch}^2}.
\end{equation}
The radial and tangential stresses are related by the momentum balance as
\begin{equation}
	\sigma_\mathrm{shell,\varphi\varphi} = -\alpha\lambda_\mathrm{ch}^3\sigma_\mathrm{shell,rr}
	\label{eq:stress-relation}
\end{equation}
with the parameter $\alpha = \frac{1}{2}\left(\frac{R_\mathrm{core}}{L_\mathrm{shell}}-1\right)$ defined by the core radius $R_\mathrm{core}$ and the shell thickness $L_\mathrm{shell}$.

Using equations (\ref{eq:boundary-condition}), (\ref{eq:pde-visc-stress}), (\ref{eq:pde-time-derivative}), and (\ref{eq:stress-relation}), we find the differential equation for the radial stress component
\begin{align}
	\frac{\mathrm{d}\sigma_\mathrm{ev}}{\mathrm{d}t} &=  E_\mathrm{core} \cdot \frac{\dot{\mathbf{E}}_\mathrm{core,\varphi\varphi}}{\lambda_\mathrm{ch}^2}\\
	&=  \frac{E_\mathrm{core}}{\tau\lambda_\mathrm{ch}^2} \operatorname{sinh}\left(\frac{\sigma_\mathrm{shell,visc,\varphi\varphi}}{\sigma_\mathrm{ref}}\right)\\
	&= -  \frac{E_\mathrm{core}}{\tau\lambda_\mathrm{ch}^2} \operatorname{sinh}\left(\frac{\alpha\lambda_\mathrm{ch}^3\sigma_\mathrm{ev}}{\sigma_\mathrm{ref}}\right).\label{eq:pde}
\end{align}
We solve the simplified differential equation in \cref{eq:pde} analytically to describe the whole time dependence with a single analytical solution
\begin{equation}
	\sigma_\mathrm{ev} = \frac{2 \sigma_\mathrm{ref}}{\alpha\lambda_\mathrm{ch}^3} \cdot \operatorname{atanh}\left(C \cdot \operatorname{exp} \left(-\frac{E_\mathrm{core}\alpha\lambda_\mathrm{ch}}{ \tau \sigma_\mathrm{ref}} t\right)\right),
	\label{eq:stress-evolution-full}
\end{equation}
where the constant $C$ can be determined from the boundary condition at time $t=0$ with $\sigma_\mathrm{ev}(t=0) = \sigma_0$.

For the calculation of the stress effect on the silicon voltage according to \cref{eq:ocv}, we approximate the deformation of the silicon particle core as purely chemical, $\mathbf{F}_\mathrm{core} = \mathbf{F}_\mathrm{core,ch}=\lambda_\mathrm{ch}\mathbf{Id}$, and we assume isotropic stress distribution inside the particle $\mathbf{P}_\mathrm{core,ev}=\lambda_\mathrm{ch}^2 \sigma_\mathrm{ev} \mathbf{Id}$.
Therefore, the impact of the stress on the voltage according to \cref{eq:ocv} simplifies to $\Delta U_\mathrm{ev}=v_\mathrm{Li}\sigma_\mathrm{ev}/F$ in the reduced model and the voltage relaxation reads
\begin{equation}
	\Delta U_\mathrm{ev} = \frac{2v_\mathrm{Li} \sigma_\mathrm{ref}}{\alpha F \lambda_\mathrm{ch}^3} \operatorname{atanh}\left(C \operatorname{exp}\left(-\frac{E_\mathrm{core}\alpha\lambda_\mathrm{ch}}{ \tau \sigma_\mathrm{ref}} t\right)\right).
\end{equation}

\noindent
To understand the origin and the regimes of the convoluted functional behavior in \cref{eq:stress-evolution-full}, we analyze the relaxation behavior in the limits of large and low stress magnitudes in section SI in the Supporting Information. Due to the importance of the long-term relaxation, here we present only the large stress limit. In the limit of large compressive stress, the differential equation (\ref{eq:pde}) simplifies to
\begin{equation}
	\frac{\mathrm{d}\sigma_\mathrm{ev}}{\mathrm{d}t} = - \frac{E_\mathrm{core}}{\tau \lambda_\mathrm{ch}^2} \cdot \left(-\frac{1}{2}\right) \operatorname{exp}\left(-\frac{\alpha\lambda_\mathrm{ch}^3\sigma_\mathrm{ev}}{\sigma_\mathrm{ref}} \right).
\end{equation}
The analytical solution for this differential equation is
\begin{equation}
	\sigma_\mathrm{ev} = \frac{\sigma_\mathrm{ref}}{\alpha\lambda_\mathrm{ch}^3} \cdot \operatorname{ln}\left(\frac{E_\mathrm{core}\alpha\lambda_\mathrm{ch}}{2 \tau \sigma_\mathrm{ref}} t + C_\mathrm{exp}\right)
\end{equation}
with the integration constant $C_\mathrm{exp}$ determined from the boundary condition.

Thus, the voltage relaxation according to the Garofalo viscosity
\begin{equation}
	\Delta U_\mathrm{ev} = \frac{v_\mathrm{Li}\sigma_\mathrm{ref}}{\alpha F \lambda_\mathrm{ch}^3} \cdot \operatorname{ln}\left(\frac{E_\mathrm{core}\alpha\lambda_\mathrm{ch}}{2 \tau \sigma_\mathrm{ref}} t + C_\mathrm{exp}\right)
\end{equation}
reveals logarithmic behavior in the large stress limit.

\subsection{Reduced Model Equations}

Complementary to our full model presented in \cref{sec:model}, we derive a reduced model with the key features in section SII in the Supporting Information. The reduced model describes the elastic stress contribution of the core at the interface between core and shell due to elastoplastic behavior of the shell $\sigma_\mathrm{ee}$ and due to viscous behavior of the shell $\sigma_\mathrm{ev}$.

The system of equations defining the reduced chemo-mechanical hysteresis model reads
\begin{align}
		\frac{\mathrm{d}\,\mathrm{SOC}}{\mathrm{d}t} &= \frac{\dot{c}_\mathrm{Li,0}}{c_\mathrm{Li,max}}  = \pm \frac{C_\mathrm{rate}}{3600} \,\frac{1}{\mathrm{s}}\label{eq:red-model-soc}\\
		\frac{\mathrm{d}\,\Delta U_\mathrm{ee}}{\mathrm{d}t} &=
		\begin{dcases}
			-E_\mathrm{shell}\frac{2v_\mathrm{Li}^2}{3F\lambda_\mathrm{ch}^7} \dot{c}_\mathrm{Li,0},& \text{if } f_\mathrm{red} < 0\\
			\frac{\alpha\sigma_\mathrm{Y,shell}v_\mathrm{Li}^2}{F\left(1+\alpha\lambda_\mathrm{ch}^3\right)^{2}}   \left|\dot{c}_\mathrm{Li,0}\right|,\quad              & \text{otherwise}
		\end{dcases}\label{eq:red-model-el}\\
		\frac{\mathrm{d}\,\Delta U_\mathrm{ev}}{\mathrm{d}t}  &= -  \frac{E_\mathrm{core}v_\mathrm{Li}}{\tau F \lambda_\mathrm{ch}^2} \operatorname{sinh}\left(\frac{\alpha\lambda_\mathrm{ch}^3 F \Delta U_\mathrm{ev}}{ \sigma_\mathrm{ref}v_\mathrm{Li}}\right)
		-\frac{E_\mathrm{core}v_\mathrm{Li}^2}{3F\lambda_\mathrm{ch}^3} \dot{c}_\mathrm{Li,0}\label{eq:red-model-visc}
\end{align}

\noindent
with the parameter $\alpha = \frac{1}{2}\left(\frac{R_\mathrm{core}}{L_\mathrm{shell}}-1\right)$ and the yield condition for plastic flow for the reduced model

\begin{equation}
		f_\mathrm{red} = -\operatorname{sgn}\left(\dot{c}_\mathrm{Li,0}\right) \left(1+\alpha\lambda_\mathrm{ch}^3\right)\frac{F \Delta U_\mathrm{ee}}{v_\mathrm{Li}\sigma_\mathrm{Y,shell}} -1 < 0.
\end{equation}

\noindent
The equations defining the reduced model describe the silicon anode voltage as $U = U_\mathrm{mean} + \Delta U_\mathrm{ee} + \Delta U_\mathrm{ev}$.
\Cref{eq:red-model-soc} states the change of SOC for lithiation ($+$) and delithiation ($-$).
The upper case in \cref{eq:red-model-el} describes the voltage evolution caused by elastic behavior of the silicon core due to elastic behavior of the shell. The lower case describes elastic core stress due to plastic behavior of the shell. The first term in \cref{eq:red-model-visc} considers the viscous shell stress relaxation. The second term considers viscous shell stress increase because of silicon volume changes.

\begin{figure}[tbp]
	\centering
	\includegraphics[width=0.47\textwidth]{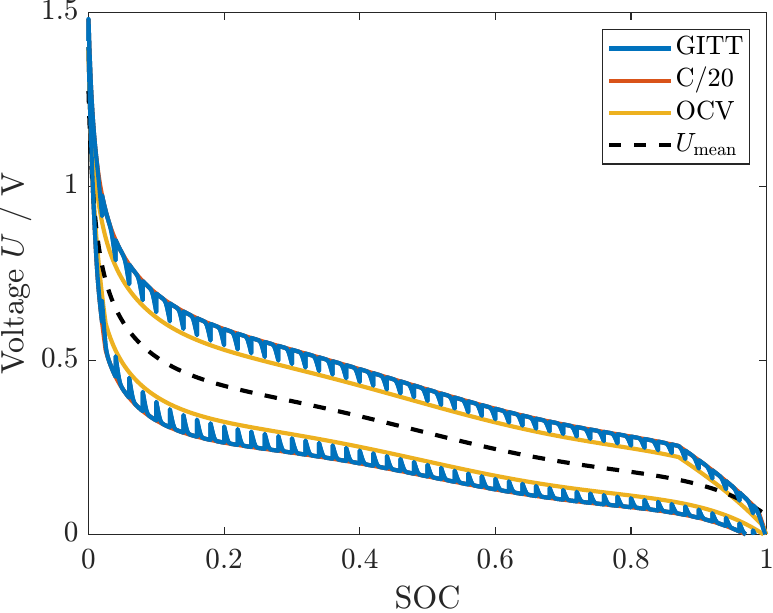}
	\caption{Voltages according to the presented reduced model during GITT, $\mathrm{C}/20$ cycling, and after $12\,\mathrm{h}$ relaxation periods. The dashed black line depicts the mean OCV measured for a silicon anode in Ref. \cite{Pan2019}.}
	\label{fig:GITT-simple}
\end{figure}

In \cref{fig:GITT-simple}, we depict the voltage profile predicted by the reduced model for a GITT procedure with (de)lithiation steps of $\Delta \mathrm{SOC}=0.02\%$ with $\mathrm{C}/20$ and relaxation periods of $3\,\mathrm{h}$. Furthermore, the figure shows the voltage during $\mathrm{C}/20$ cycling and after $12\,\mathrm{h}$ relaxation periods. The dashed black line depicts the fitted mean OCV curve $U_\mathrm{mean}$ between the measured lithiation and delithiation voltage after $3\,\mathrm{h}$ rest period for a pure silicon anode from Ref. \cite{Pan2019} used as true OCV curve for the simulations.

\begin{figure*}[htbp]
	\centering
	\includegraphics[height=0.37\textwidth]{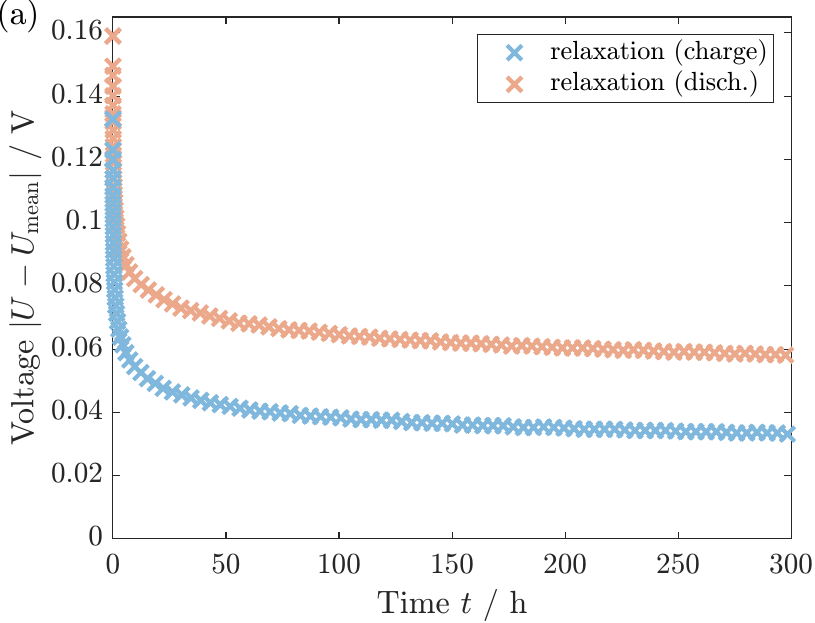}
	\hspace{0.5cm}
	\includegraphics[height=0.37\textwidth]{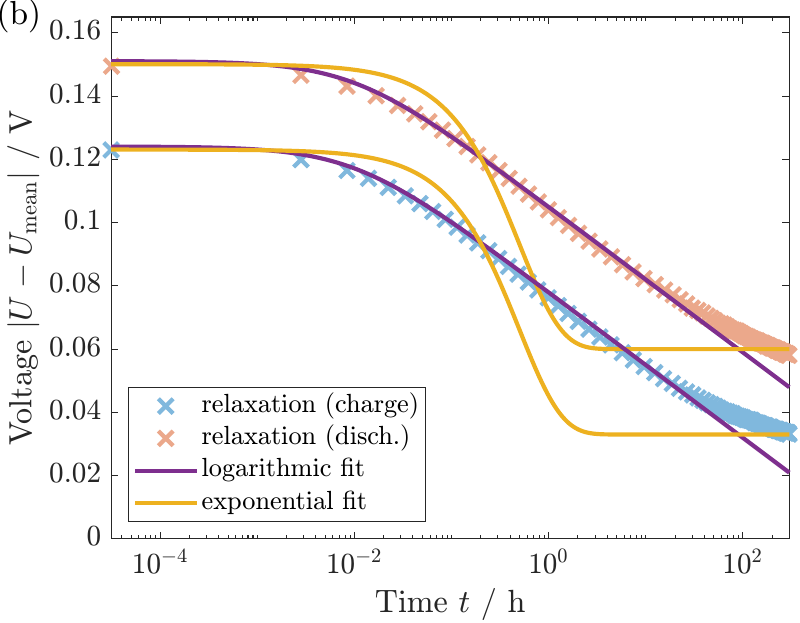}
	\caption{(a) Experimental voltage relaxation of silicon at $\mathrm{SOC}=0.3$ over 300 hours after a charge and discharge period (protocol SIII\,A) \cite{Wycisk2023}. (b) The semi-logarithmic plot unveils the logarithmic voltage relaxation behavior.}
	\label{fig:relaxation-exp}
\end{figure*}

\section{Computational \& Experimental Details}
\label{sec:experimental}

\subsection{Simulation Setup}
Our simulations describe the behavior of a silicon nanoparticle anode with a single-particle model. We implement our model in MATLAB using a finite-difference approach by discretizing the radial dimension. To solve the set of differential equations (\ref{eq:dot-concentration}), (\ref{eq:nabla-stress}), (\ref{eq:plastic-flow}), (\ref{eq:nabla-stress-sei}), and (\ref{eq:plastic-flow-sei}), we use the solver ode15i. The variables inside the silicon core are the concentration of lithium $c_{\mathrm{Li},0}$, the deformed radius $r_\mathrm{core}$, and the radial component of the plastic deformation $\mathbf{F}_{\mathrm{core,pl,rr}}$ of each silicon core element. The variables inside the SEI shell are the deformed radius $r_\mathrm{shell}$ and the radial component of the plastic deformation $\mathbf{F}_{\mathrm{shell,pl,rr}}$ of each SEI shell element.

\subsection{Material Parameters}
We adopt the parameters from our previous publication \cite{Koebbing2023Voltage} and adapt where necessary. Particularly, we consider the stiff, inorganic SEI shell with Young's modulus of $E_\mathrm{shell}=100\,\mathrm{GPa}$ compatible with experiments \cite{Shin2015, Chai2021}. The viscosity of the inner SEI shell is considered as a fit value and may range from $\eta=10^7\,\mathrm{Pa\,s}$ for a highly viscous polymer \cite{Edgeworth1984} to $\eta=10^{15}\,\mathrm{Pa\,s}$ for silicon oxide \cite{Sutardja1989, Senez1994, Ojovan2008}.

\subsection{Experimental Setup}
The experiments analyzed in this study have been performed and published by Wycisk et al. \cite{Wycisk2023} at Mercedes following discussions with the authors of this manuscript. The publication discusses full-cell voltage measurements with an NMC811 cathode and anodes with varying contents of silicon active material. Here, we constrain solely to the experimental results discussing anodes with pure silicon active material. The silicon anode consists of silicon nanoparticles attached to a conductive carbon network discussed as "silicon-carbon composite granules" in Ref. \cite{Schwan2020}. 
We summarize the experimental and our simulation protocols in section SIII in the Supporting Information but refer to the experimental publication for the experimental details \cite{Wycisk2023}.

Throughout this manuscript, we consider voltages from the anode perspective and calculate voltage differences to the mean OCV, $U-U_\mathrm{mean}$. For comparison, the voltage difference for the full-cell measurements is calculated as $U-U_\mathrm{mean} = - (U_\mathrm{full}-U_\mathrm{full,mean})$.

\begin{figure*}[htbp]
	\centering
	\includegraphics[height=0.37\textwidth]{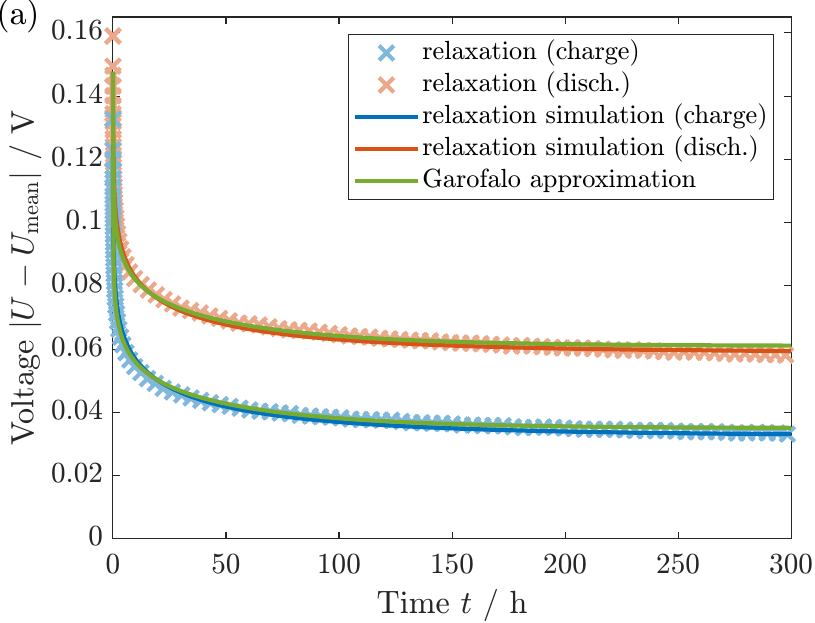}
	\hspace{0.5cm}
	\includegraphics[height=0.37\textwidth]{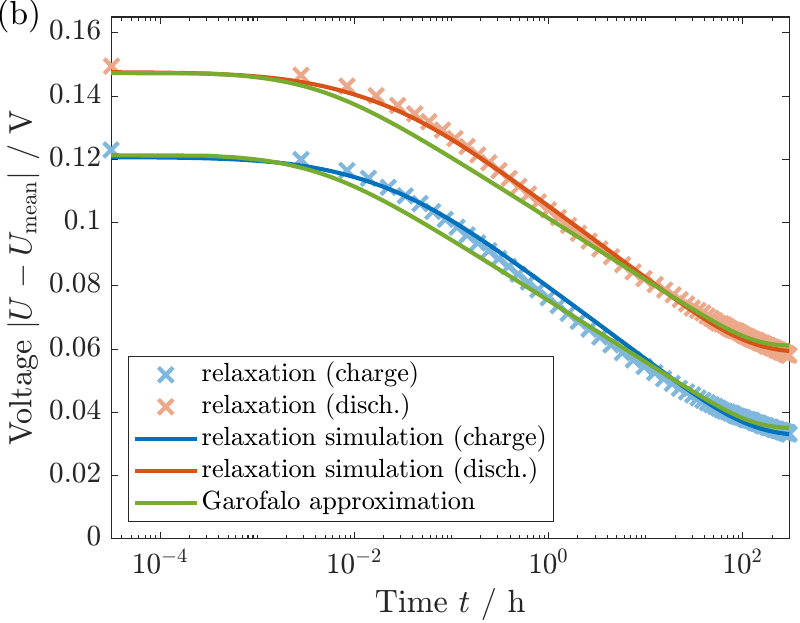}
	\caption{(a) Voltage relaxation of silicon at $\mathrm{SOC}=0.3$ over 300 hours (protocol SIII\,A). Comparison of simulation, experiment \cite{Wycisk2023}, and the analytical Garofalo approximation. (b) The semi-logarithmic plot shows agreement of the various curves.}
	\label{fig:relaxation-pde}
\end{figure*}

\section{Results and Discussion}
\label{sec:results}

\subsection{Experimental Results: Logarithmic Voltage Relaxation}

First, we analyze the long-time relaxation experiment performed by Wycisk et al. \cite{Wycisk2023} following the protocol described in section SIII\,A in the Supporting Information. In \cref{fig:relaxation-exp}, we depict the voltage relaxation at the same SOC measured once in charge and once in discharge direction.

Interestingly, the authors of Ref. \cite{Wycisk2023} find that even after $300\,\mathrm{h}$ of rest, the voltage depicted in \cref{fig:relaxation-exp}(a) is not completely relaxed. Therefore, the true OCV value deviates from the relaxed voltage after $300\,\mathrm{h}$ and strongly deviates from standard GITT measurements with only a few hours of voltage relaxation. The authors of Ref. \cite{Wycisk2023} exclude degradation or self-discharge due to the similar voltage relaxation profiles after the charge and discharge period.
However, the mean value of the relaxed voltage after $300\,\mathrm{h}$ varies from the mean OCV measured with GITT for $\mathrm{C}/20$ and $12\,\mathrm{h}$ rest period due to deviations of the experimental cells.

Here, we investigate the voltage relaxation profile in detail again. In \cref{fig:relaxation-exp}(b), we show the voltage relaxation over time as a semi-logarithmic plot. 
Apparently, the voltage relaxation profile does not follow a typical exponential relaxation behavior as illustrated in yellow.
We identify a linear regime in the semi-logarithmic plot and fit a logarithmic function to the experimental data. The logarithmic fit agrees with the experimental data in a wide range of times $t < 20\,\mathrm{h}$. Only for times larger than $20\,\mathrm{h}$, the voltage relaxation slightly diminishes leaving the logarithmic regime. This is expected as logarithmic behavior would diverge for large times. The logarithmic voltage relaxation found in the experiment agrees with the experimentally observed voltage relaxation of silicon thin-film electrodes in Ref. \cite{Sethuraman2013}. It is in stark contrast to diffusion limitation with exponential long-term behavior and supports the theory of a mechanical origin.

\subsection{Simulation Results: Slow Voltage Relaxation}

As discussed in \cref{sec:model}, the silicon OCV hysteresis results from elastoplastic stress generated by the shell, and the enlarged voltage hysteresis during cycling results from viscous shell stress acting on the particle core \cite{Koebbing2023Voltage}.
A simple Newtonian viscosity model, $\sigma_\mathrm{shell} = \eta_\mathrm{shell} \dot{\mathbf{E}}_\mathrm{shell}$, with constant viscosity $\eta_\mathrm{shell}$ would imply exponential voltage relaxation behavior during rest contrasting the experimental observations. Due to the large stresses inside the SEI shell, the Newtonian model is not suitable for describing the viscous behavior. Instead, for large stresses, the strain rate is known to depend exponentially on the stress, leading to a logarithmic stress relaxation behavior. Therefore, we use the established Garofalo law given in \cref{eq:Garofalo-law} to describe both regimes.

Using the Garofalo model, \cref{fig:relaxation-pde} depicts our simulation results in comparison to the experimental data. The parameters are given in the Supporting Information in Table S1. 
We shift our simulations to match the observed voltage after relaxation.
The simulations reproduce the voltage relaxation profiles after the charge and discharge period. In particular, the simulation using the Garofalo law describes both the logarithmic relaxation regime as well as the decreasing relaxation after $20\,\mathrm{h}$. The agreement confirms the explanation of the silicon voltage hysteresis by a visco-elastoplastic SEI behavior.

To validate our simulation results, \cref{fig:relaxation-pde} compares our simulation and the experiment to the analytical approximation presented in \cref{sec:analytical-solution}. The analytical approximation for the voltage relaxation with Garofalo law viscosity reveals a similar logarithmic relaxation regime followed by a slowed relaxation. Thus, the specific trends observed for our simulation and the analytical approximation agree while the actual values deviate slightly. Nevertheless, as the analytical approach relies on several assumptions and approximations, the similarity of the voltage profile supports our simulation results.

\subsection{OCV and Cycling Voltage Hysteresis}

\begin{figure}[tbp]
	\centering
	\includegraphics[width=0.47\textwidth]{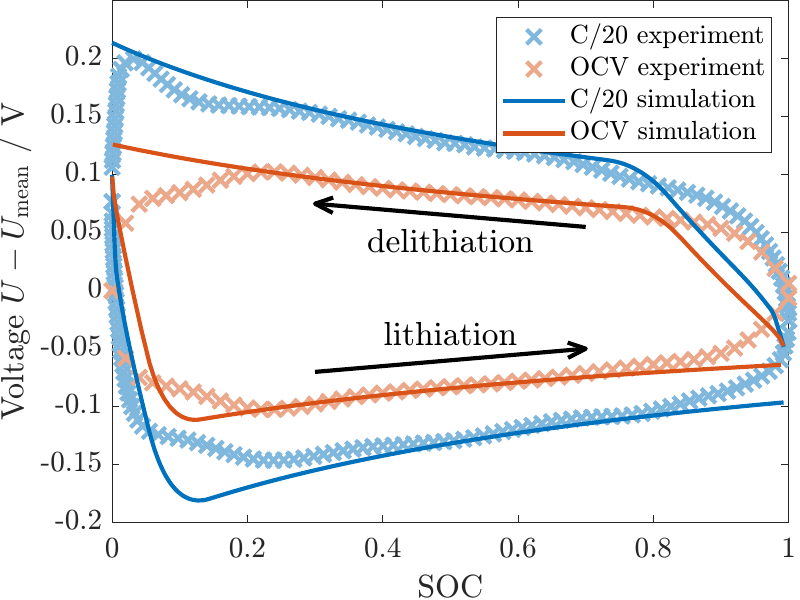}
	\caption{$\mathrm{C}/20$ and open-circuit voltage hysteresis after $12\,\mathrm{h}$ relaxation in simulation and experiment (protocol SIII\,B) \cite{Wycisk2023}.}
	\label{fig:hysteresis}
\end{figure}

Silicon anodes are generally known to show a significant voltage hysteresis. In \cref{fig:hysteresis}, we depict the experimental OCV hysteresis after long relaxation and the enlarged voltage hysteresis during slow cycling \cite{Wycisk2023}. We describe the protocol in SIII\,B in the Supporting Information. To check the consistency of our model with the experimental voltage hysteresis, \cref{fig:hysteresis} shows the simulation of the anode voltage during slow cycling and the OCV after long relaxation depending on the SOC for the parameters obtained from the voltage relaxation behavior.
The illustrated voltages describe the influence on the silicon anode voltage. Hence, the voltage decreases during lithiation due to compressive stress and increases during delithiation due to tensile stress.
The simulation results in \cref{fig:hysteresis} reveal a significant OCV hysteresis resulting from the elastoplastic contribution. Furthermore, the simulation shows an enlarged hysteresis during cycling caused by viscous stress.

The comparison of the cycling and relaxed voltages reveals a good agreement between simulation and experiment in a wide SOC regime. However, our simulation and the experiment deviate slightly at both extremes, $\mathrm{SOC} <0.2$ and $\mathrm{SOC}>0.8$. This disagreement results at least partially from the determination of the true, stress-free OCV curve as the mean between lithiation and delithiation OCV. At very high SOC, the compressive stress during lithiation is fully developed, while the tensile stress during the following delithiation has to build up gradually after the change of direction. Analogously, the tensile stress during delithiation is fully developed, while the compressive stress during the following lithiation has to build up gradually after the change of direction at low SOC. Therefore, the mean value between the lithiation and delithiation OCV at both extremes is not stress-free. Its consideration as true, stress-free OCV in the simulation leads to an apparent deviation. In the Supporting Information, we discuss a corrected OCV curve assuming a constant hysteresis size in the extreme SOC regimes. Fig. S3  reveals a better agreement between simulation and experiment in the extreme SOC regimes.

In our previous publication \cite{Koebbing2023Voltage}, we compared our simulation to the GITT measurement performed for a silicon half cell by Pan et al. \cite{Pan2019, Pan2020}. The cells differ significantly from the cells investigated by Wycisk et al. \cite{Wycisk2023} due to a presumably different silicon raw material and electrolyte composition. Nevertheless, we compare our new model and the parameters obtained from the voltage relaxation \cite{Wycisk2023} to the GITT measurement \cite{Pan2019,Pan2020} in section SV in the Supporting Information. Fig. S4 shows a reasonable match of simulation and experiment considering the full GITT procedure as well as a single GITT pulse. The agreement confirms the applicability of our chemo-mechanical model to GITT measurements with different cells.

\subsection{C-Rate Dependence of Voltage Hysteresis}

\begin{figure}[tbp]
	\centering
	\includegraphics[width=0.47\textwidth]{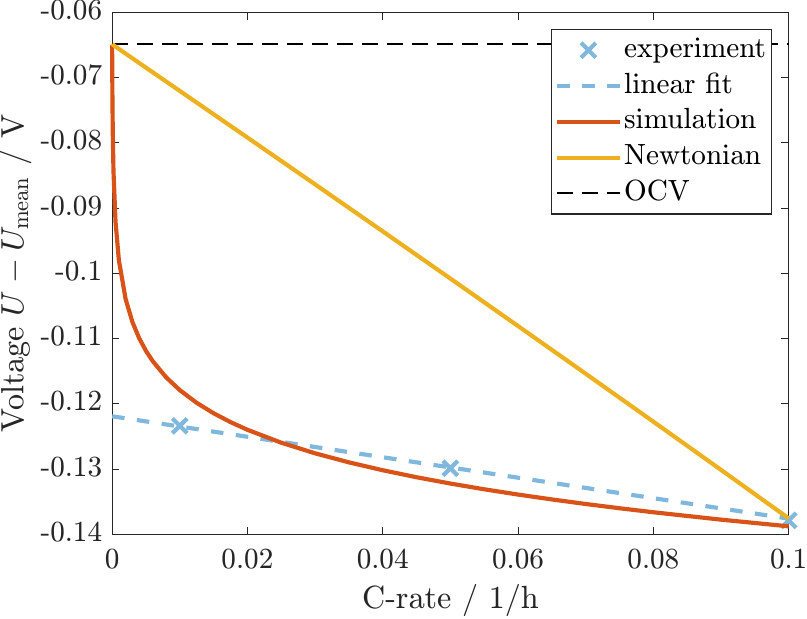}
	\caption{C-rate dependence of voltage hysteresis at $\mathrm{SOC}=0.5$ in simulation and experiment (protocol SIII\,C) \cite{Wycisk2023}.}
	\label{fig:hysteresis-C-rates}
\end{figure}

The experimental data obtained by Wycisk et al. \cite{Wycisk2023} also cover the C-rate dependence of the voltage difference between the cycling voltage and the relaxed voltage after $12\,\mathrm{h}$ at $\mathrm{SOC}=0.5$ following the protocol given in SIII\,C in the Supporting Information. As displayed in \cref{fig:hysteresis-C-rates}, the data reveal a linear dependence of the voltage on the C-rate. However, extrapolating this linear dependence to zero current results in a significant voltage offset compared to the OCV after infinite relaxation time. This offset would imply an enlarged hysteresis even for infinitely slow cycling, which is unexpected. Therefore, the authors conclude that the voltage will depart from the linear trend at particularly low C-rates.

The Newtonian viscosity model has a linear relation between the strain rate and the viscous stress. Hence, the size of the additional voltage hysteresis is linearly dependent on the C-rate as illustrated in yellow in \cref{fig:hysteresis-C-rates}. However, the Newtonian model explains no voltage offset, and the slope disagrees with the experiment when matching the hysteresis size at $\mathrm{C}/10$.

In comparison to the experimental and the Newtonian C-rate dependence, \cref{fig:hysteresis-C-rates} also depicts the simulated C-rate dependence. The inverse hyperbolic sine in \cref{eq:Garofalo-law} determines the C-rate dependence of the viscous stress and, consequently, the C-rate dependence of the additional voltage hysteresis during cycling. Thus, the simulation reveals a non-linear dependence of the voltage on the current. Nonetheless, after a swift increase of the voltage at current rates smaller $\mathrm{C}/100$, the increase slows down, approaching an almost linear trend with small curvature. Although the three experimental data points follow the linear trend exactly, we assume that our simulation is in reasonable agreement with the experiment and additionally describes the transition to vanishing voltage at zero current. We expect that more experimental data points particularly at low C-rates might indicate a curvature and deviation from the linear trend.

\subsection{Voltage Transition Profiles}

Another interesting behavior is the silicon anode voltage profile of transitions between cycling and rest periods. In the following, we discuss the features of different transitions and compare our simulation to the experimental data from Ref. \cite{Wycisk2023} wherever possible.

\begin{figure}[tbp]
	\centering
	\includegraphics[width=0.47\textwidth]{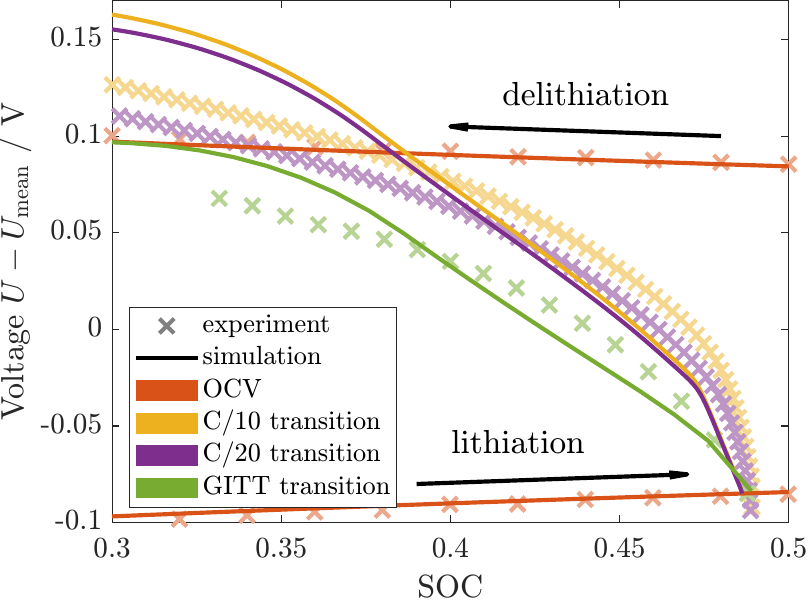}
	\caption{Voltage transition from lithiation to delithiation in simulation and experiment (protocol SIII\,D) \cite{Wycisk2023}.}
	\label{fig:hysteresis-transition}
\end{figure}

First, we investigate the transition profile between lithiation and delithiation according to protocol SIII\,D in the Supporting Information. In \cref{fig:hysteresis-transition}, we show the delithiation with either $\mathrm{C}/10$, $\mathrm{C}/20$, or GITT procedure after a continuous lithiation and rest period. For reference, the figure also includes the simulated and measured lithiation and delithiation OCV curves from \cref{fig:hysteresis}, which almost coincide in the depicted regime $0.3 < \mathrm{SOC} < 0.5$. All experimental data \cite{Wycisk2023} reveal a smooth transition between the lithiation and delithiation voltage. The slope of the voltage profiles is large directly after the change of direction and slows down gradually when approaching the delithiation voltage.

The numerical results are depicted in \cref{fig:hysteresis-transition} compared to the experiment. When switching the current direction from lithiation to delithiation, the simulated voltage profiles for $\mathrm{C}/10$ (yellow) and $\mathrm{C}/20$ (purple)  currents reveal three regimes. Immediately after the change of direction, the voltage shows a steep increase for a small span of $\Delta \mathrm{SOC} \approx 0.01$ attributed to the rapid build-up of viscous stress. Afterward, for a range of $\Delta \mathrm{SOC} \approx 0.1$, a constant, moderate voltage slope demonstrates the decrease of compressive elastic stress and the subsequent increase in tensile elastic stress. In the third regime, the slope slows down, and the voltage approaches a maximum value when reaching the yield criterion for plasticity. The higher current $\mathrm{C}/10$ shows a slightly faster voltage transition compared to the lower current $\mathrm{C}/20$. For the GITT transition curve (green), the relaxation of viscous stress during the rest periods suppresses the viscous regime after the change of direction. Contrary to the simulation, the experimental curves do not reveal clearly defined regimes but are in line with the general trend of a rapid voltage increase after the change of direction followed by an attenuated transition to the delithiation voltage curve. The much smoother experimental results compared to our simulation are expected as we consider only a single-particle model but the detailed features average out in the multi-particle experiment. Thus, we conclude that our simulation result agrees reasonably with the experimental measurement.

In the Supporting Information in section SVI, we evaluate the behavior of an interrupted lithiation pulse for different C-rates and at different SOC values. All voltage profiles in Figures S5 and S6 show a steep slope at the beginning of the pulses, revealing the increase in viscous stress followed by a slower convergence to the lithiation voltage, indicating elastoplastic behavior. The similar voltage profiles for different C-rates indicate that the voltage transition needs a certain amount of charge throughput or SOC change $\Delta \mathrm{SOC}$ in accordance with the experimental results from Ref. \cite{Wycisk2023} for a blended graphite-silicon anode.
Additionally, the voltage profiles at different SOC values in Fig. S6 show that the general trends of the chemo-mechanical simulation agree with the ones of the experiment. However, all experimental curves show an overshoot instead of a smooth convergence to the lithiation voltage, which is not visible in our simulations. In terms of mechanics, this overshoot might result from a thixotropic behavior of the SEI shell as discussed in the Supporting Information.

Another voltage hysteresis effect measured for silicon anodes is a pronounced relaxation during rest observed for higher applied currents \cite{Wycisk2023, Durdel2023}. Higher C-rates show an increased voltage hysteresis during cycling in agreement with our viscosity model. However, this dependence surprisingly inverts after relaxation.
This phenomenon is not captured in our chemo-mechanical single-particle model. Therefore, we support the interpretation as a multi-particle effect \cite{Wycisk2023} and add a mechanical explanation.
For fast charging, the silicon particles inside the anode will lithiate more inhomogeneously, causing enhanced plastic flow of the shell around particles with a higher lithiation level. During the subsequent rest period, the silicon particles with initially higher lithiation degrees delithiate slightly. The shrinkage of those particles reduces the remaining compressive stress, while the stress in the particles with initially lower lithiation levels can not exceed the yield stress for plastic flow. Hence, this multi-particle effect can reduce the mean stress hysteresis inside the silicon anode and, consequently, the voltage hysteresis after relaxation.

\begin{figure*}[htbp]
	\centering
	\includegraphics[height=0.35\textwidth]{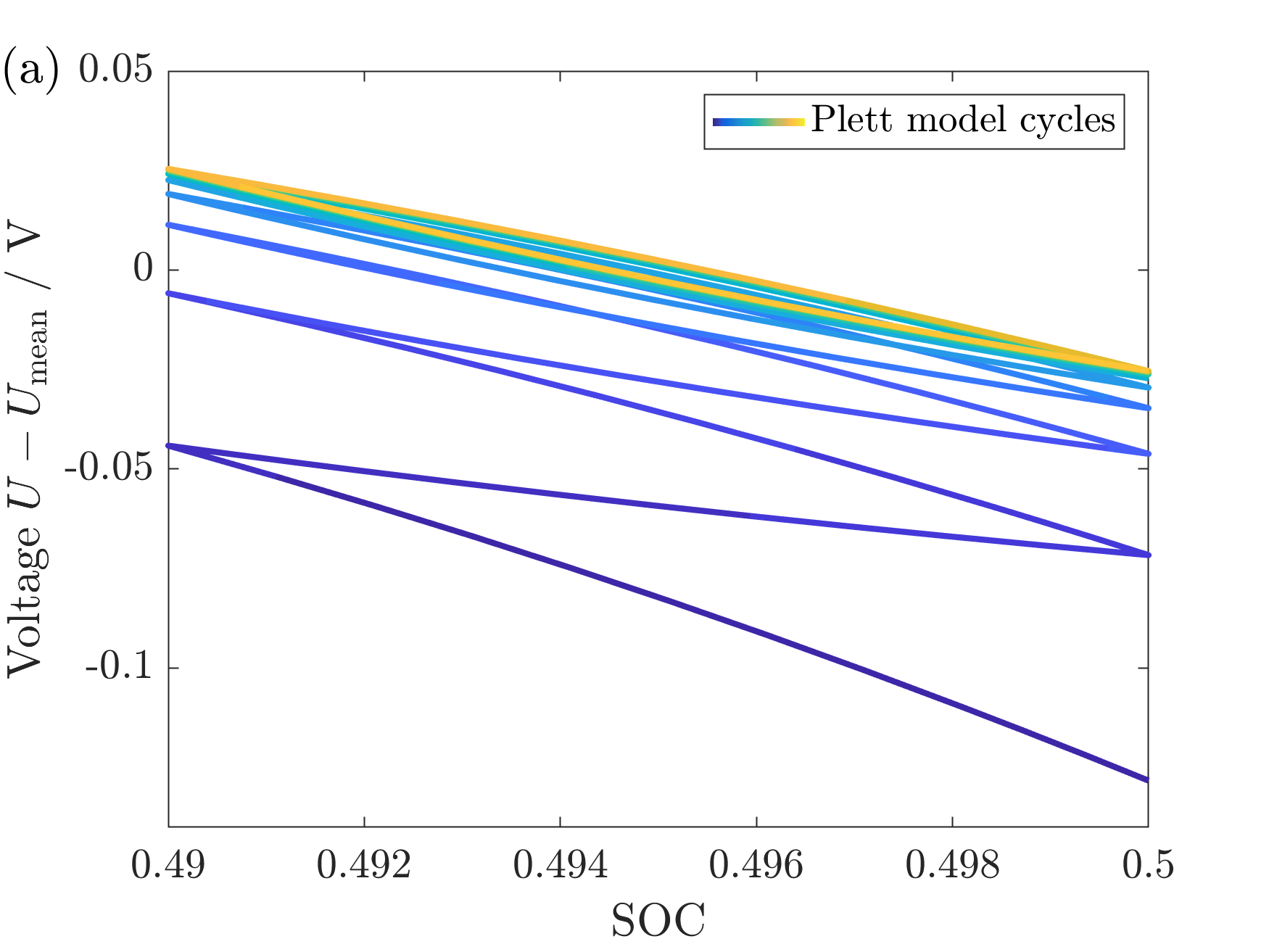}
	\hspace{0.5cm}
	\includegraphics[height=0.35\textwidth]{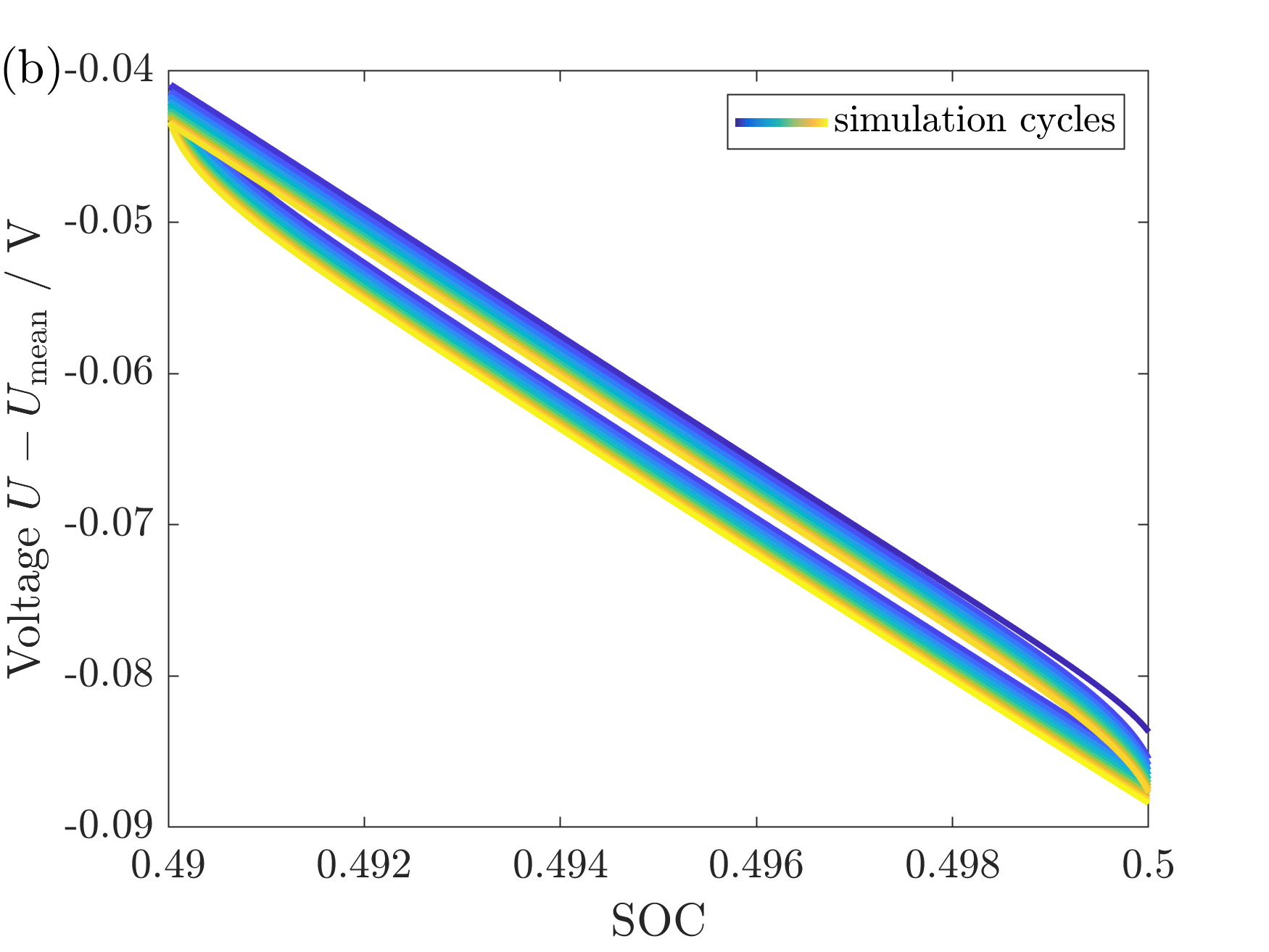}
	\caption{Voltage for alternating lithiation and delithiation pulses with $\Delta \mathrm{SOC}= 1\%$ (protocol SIII\,F) for (a) the phenomenological Plett model and (b) our chemo-mechanical simulation.}
	\label{fig:pulses-SOC-neutral}
\end{figure*}

Finally, we estimate the voltage transition behavior for alternating short lithiation and delithiation pulses following protocol SIII\,F in the Supporting Information. 
The silicon voltage hysteresis is often described empirically with the Plett model presented in section SVII in the Supporting Information \cite{Plett2004, Graells2020, Wycisk2022}. In \cref{fig:pulses-SOC-neutral}(a), we depict the behavior for alternating pulses with $\Delta \mathrm{SOC} = 0.01$ predicted by the empirical Plett model with the parameters adjusted to fit the experimental voltage hysteresis. The Plett model does not reveal a constant hysteresis behavior during $10$ subsequent cycles but rather approaches the mean OCV within the first cycles and then describes a hysteresis around it. Additionally, the Plett model is not able to account for a relaxation phase without a change in SOC.
In contrast, \cref{fig:pulses-SOC-neutral}(b) shows the simulation of alternating pulses, which reveal a permanent hysteresis during $10$ subsequent cycles. Only the very first pulse initially shows a slightly different behavior with an enlarged hysteresis size because of a different stress state in the initial situation after the $12\,\mathrm{h}$ relaxation period. 
We know that experiments show a permanent hysteresis behavior upon alternate lithiation and delithiation pulses in line with our simulation result. Thus, we conclude that our chemo-mechanical core-shell model outperforms the empirical Plett model in the description of voltage hysteresis phenomena.

\section{Conclusions}
\label{sec:conclusion}

Detailed analysis of the silicon voltage hysteresis experiments performed by Wycisk et al. \cite{Wycisk2023} reveals a slow, non-exponential voltage relaxation. We identify a logarithmic voltage relaxation for a wide range of times and a transition to exponential relaxation for larger times due to the divergence of the logarithmic behavior. With a chemo-mechanical core-shell model, we have illustrated that the visco-elastoplastic SEI shell behavior following the Garofalo law or inverse hyperbolic sine law for viscosity can accurately describe the voltage relaxation of a silicon anode over the whole time span. Our simulations also reproduce the observed voltage hysteresis and GITT measurement with the parameters obtained from the relaxation experiment.

Additionally, the Garofalo viscosity model can approach the experimentally observed C-rate dependence of the cycling voltage hysteresis. The inverse hyperbolic sine behaves approximately linear in a wide span of C-rates but shows a kink and reveals vanishing additional voltage hysteresis at zero current. Therefore, the Garofalo law viscosity model fits much better to the C-rate dependence than Newtonian viscosity, which reveals a proportional relation between the voltage and the applied C-rate.

With a focus on the voltage transition behavior between lithiation and delithiation, the presented chemo-mechanical model can adequately describe the general trends of an initially fast voltage transition followed by an attenuated convergence to the delithiation voltage curve. The interplay of viscous, elastic, and plastic contributions to the simulated voltage explains this voltage profile. Furthermore, our model reasonably describes the lithiation behavior after a rest period.
Thus, our chemo-mechanical core-shell model outperforms the empirical Plett model regarding physical understanding as well as the description of the various features of the hysteresis phenomenon.

The overall accordance of our simulations to experimental results supports our chemo-mechanical explanation of the voltage hysteresis presented initially in Ref. \cite{Koebbing2023Voltage}. The description of the viscous behavior using the Garofalo law is more suitable than linear Newtonian viscosity because of the large stresses reached inside the SEI shell. In conclusion, we have demonstrated that our physical model presents a consistent picture of the various features of the silicon voltage hysteresis phenomenon.

\vspace{0.1cm}
\begin{acknowledgments}
Lukas Köbbing gratefully acknowledges funding and support by the European Union’s Horizon Europe within the research initiative Battery 2030+ via the OPINCHARGE project under the grant agreement number 101104032 and
by the German Research Foundation (DFG) within the research training group SiMET under project number 281041241/GRK2218. The authors appreciate fruitful discussions with Dominik Wycisk, Felix Schwab, Martin Werres, Raphael Schoof, and Arnulf Latz.
\end{acknowledgments}

\bibliography{refs}

\begin{thebibliography}{54}%
\makeatletter
\providecommand \@ifxundefined [1]{%
 \@ifx{#1\undefined}
}%
\providecommand \@ifnum [1]{%
 \ifnum #1\expandafter \@firstoftwo
 \else \expandafter \@secondoftwo
 \fi
}%
\providecommand \@ifx [1]{%
 \ifx #1\expandafter \@firstoftwo
 \else \expandafter \@secondoftwo
 \fi
}%
\providecommand \natexlab [1]{#1}%
\providecommand \enquote  [1]{``#1''}%
\providecommand \bibnamefont  [1]{#1}%
\providecommand \bibfnamefont [1]{#1}%
\providecommand \citenamefont [1]{#1}%
\providecommand \href@noop [0]{\@secondoftwo}%
\providecommand \href [0]{\begingroup \@sanitize@url \@href}%
\providecommand \@href[1]{\@@startlink{#1}\@@href}%
\providecommand \@@href[1]{\endgroup#1\@@endlink}%
\providecommand \@sanitize@url [0]{\catcode `\\12\catcode `\$12\catcode
  `\&12\catcode `\#12\catcode `\^12\catcode `\_12\catcode `\%12\relax}%
\providecommand \@@startlink[1]{}%
\providecommand \@@endlink[0]{}%
\providecommand \url  [0]{\begingroup\@sanitize@url \@url }%
\providecommand \@url [1]{\endgroup\@href {#1}{\urlprefix }}%
\providecommand \urlprefix  [0]{URL }%
\providecommand \Eprint [0]{\href }%
\providecommand \doibase [0]{https://doi.org/}%
\providecommand \selectlanguage [0]{\@gobble}%
\providecommand \bibinfo  [0]{\@secondoftwo}%
\providecommand \bibfield  [0]{\@secondoftwo}%
\providecommand \translation [1]{[#1]}%
\providecommand \BibitemOpen [0]{}%
\providecommand \bibitemStop [0]{}%
\providecommand \bibitemNoStop [0]{.\EOS\space}%
\providecommand \EOS [0]{\spacefactor3000\relax}%
\providecommand \BibitemShut  [1]{\csname bibitem#1\endcsname}%
\let\auto@bib@innerbib\@empty
\bibitem [{\citenamefont {Sun}\ \emph {et~al.}(2022)\citenamefont {Sun},
  \citenamefont {Liu}, \citenamefont {Shao}, \citenamefont {Wu}, \citenamefont
  {Jiang},\ and\ \citenamefont {Jin}}]{Sun2022}%
  \BibitemOpen
  \bibfield  {author} {\bibinfo {author} {\bibfnamefont {L.}~\bibnamefont
  {Sun}}, \bibinfo {author} {\bibfnamefont {Y.}~\bibnamefont {Liu}}, \bibinfo
  {author} {\bibfnamefont {R.}~\bibnamefont {Shao}}, \bibinfo {author}
  {\bibfnamefont {J.}~\bibnamefont {Wu}}, \bibinfo {author} {\bibfnamefont
  {R.}~\bibnamefont {Jiang}},\ and\ \bibinfo {author} {\bibfnamefont
  {Z.}~\bibnamefont {Jin}},\ }\bibfield  {title} {\bibinfo {title} {{Recent
  progress and future perspective on practical silicon anode-based lithium ion
  batteries}},\ }\href {https://doi.org/10.1016/j.ensm.2022.01.042} {\bibfield
  {journal} {\bibinfo  {journal} {Energy Storage Materials}\ }\textbf {\bibinfo
  {volume} {46}},\ \bibinfo {pages} {482} (\bibinfo {year} {2022})}\BibitemShut
  {NoStop}%
\bibitem [{\citenamefont {Zuo}\ \emph {et~al.}(2017)\citenamefont {Zuo},
  \citenamefont {Zhu}, \citenamefont {M{\"{u}}ller-Buschbaum},\ and\
  \citenamefont {Cheng}}]{Zuo2017}%
  \BibitemOpen
  \bibfield  {author} {\bibinfo {author} {\bibfnamefont {X.}~\bibnamefont
  {Zuo}}, \bibinfo {author} {\bibfnamefont {J.}~\bibnamefont {Zhu}}, \bibinfo
  {author} {\bibfnamefont {P.}~\bibnamefont {M{\"{u}}ller-Buschbaum}},\ and\
  \bibinfo {author} {\bibfnamefont {Y.-J.}\ \bibnamefont {Cheng}},\ }\bibfield
  {title} {\bibinfo {title} {{Silicon based lithium-ion battery anodes: A
  chronicle perspective review}},\ }\href
  {https://doi.org/10.1016/j.nanoen.2016.11.013} {\bibfield  {journal}
  {\bibinfo  {journal} {Nano Energy}\ }\textbf {\bibinfo {volume} {31}},\
  \bibinfo {pages} {113} (\bibinfo {year} {2017})}\BibitemShut {NoStop}%
\bibitem [{\citenamefont {Feng}\ \emph {et~al.}(2018)\citenamefont {Feng},
  \citenamefont {Li}, \citenamefont {Liu}, \citenamefont {Kashkooli},
  \citenamefont {Xiao}, \citenamefont {Cai},\ and\ \citenamefont
  {Chen}}]{Feng2018}%
  \BibitemOpen
  \bibfield  {author} {\bibinfo {author} {\bibfnamefont {K.}~\bibnamefont
  {Feng}}, \bibinfo {author} {\bibfnamefont {M.}~\bibnamefont {Li}}, \bibinfo
  {author} {\bibfnamefont {W.}~\bibnamefont {Liu}}, \bibinfo {author}
  {\bibfnamefont {A.~G.}\ \bibnamefont {Kashkooli}}, \bibinfo {author}
  {\bibfnamefont {X.}~\bibnamefont {Xiao}}, \bibinfo {author} {\bibfnamefont
  {M.}~\bibnamefont {Cai}},\ and\ \bibinfo {author} {\bibfnamefont
  {Z.}~\bibnamefont {Chen}},\ }\bibfield  {title} {\bibinfo {title}
  {{Silicon‐Based Anodes for Lithium‐Ion Batteries: From Fundamentals to
  Practical Applications}},\ }\bibfield  {journal} {\bibinfo  {journal}
  {Small}\ }\textbf {\bibinfo {volume} {14}},\ \href
  {https://doi.org/10.1002/smll.201702737} {10.1002/smll.201702737} (\bibinfo
  {year} {2018})\BibitemShut {NoStop}%
\bibitem [{\citenamefont {Beaulieu}\ \emph {et~al.}(2001)\citenamefont
  {Beaulieu}, \citenamefont {Eberman}, \citenamefont {Turner}, \citenamefont
  {Krause},\ and\ \citenamefont {Dahn}}]{Beaulieu2001}%
  \BibitemOpen
  \bibfield  {author} {\bibinfo {author} {\bibfnamefont {L.~Y.}\ \bibnamefont
  {Beaulieu}}, \bibinfo {author} {\bibfnamefont {K.~W.}\ \bibnamefont
  {Eberman}}, \bibinfo {author} {\bibfnamefont {R.~L.}\ \bibnamefont {Turner}},
  \bibinfo {author} {\bibfnamefont {L.~J.}\ \bibnamefont {Krause}},\ and\
  \bibinfo {author} {\bibfnamefont {J.~R.}\ \bibnamefont {Dahn}},\ }\bibfield
  {title} {\bibinfo {title} {{Colossal Reversible Volume Changes in Lithium
  Alloys}},\ }\href {https://doi.org/10.1149/1.1388178} {\bibfield  {journal}
  {\bibinfo  {journal} {Electrochemical and Solid-State Letters}\ }\textbf
  {\bibinfo {volume} {4}},\ \bibinfo {pages} {A137} (\bibinfo {year}
  {2001})}\BibitemShut {NoStop}%
\bibitem [{\citenamefont {Liu}\ \emph {et~al.}(2012)\citenamefont {Liu},
  \citenamefont {Zhong}, \citenamefont {Huang}, \citenamefont {Mao},
  \citenamefont {Zhu},\ and\ \citenamefont {Huang}}]{Liu2012}%
  \BibitemOpen
  \bibfield  {author} {\bibinfo {author} {\bibfnamefont {X.~H.}\ \bibnamefont
  {Liu}}, \bibinfo {author} {\bibfnamefont {L.}~\bibnamefont {Zhong}}, \bibinfo
  {author} {\bibfnamefont {S.}~\bibnamefont {Huang}}, \bibinfo {author}
  {\bibfnamefont {S.~X.}\ \bibnamefont {Mao}}, \bibinfo {author} {\bibfnamefont
  {T.}~\bibnamefont {Zhu}},\ and\ \bibinfo {author} {\bibfnamefont {J.~Y.}\
  \bibnamefont {Huang}},\ }\bibfield  {title} {\bibinfo {title}
  {{Size-Dependent Fracture of Silicon Nanoparticles During Lithiation}},\
  }\href {https://doi.org/10.1021/nn204476h} {\bibfield  {journal} {\bibinfo
  {journal} {ACS Nano}\ }\textbf {\bibinfo {volume} {6}},\ \bibinfo {pages}
  {1522} (\bibinfo {year} {2012})}\BibitemShut {NoStop}%
\bibitem [{\citenamefont {Wetjen}\ \emph {et~al.}(2018)\citenamefont {Wetjen},
  \citenamefont {Solchenbach}, \citenamefont {Pritzl}, \citenamefont {Hou},
  \citenamefont {Tileli},\ and\ \citenamefont {Gasteiger}}]{Wetjen2018}%
  \BibitemOpen
  \bibfield  {author} {\bibinfo {author} {\bibfnamefont {M.}~\bibnamefont
  {Wetjen}}, \bibinfo {author} {\bibfnamefont {S.}~\bibnamefont {Solchenbach}},
  \bibinfo {author} {\bibfnamefont {D.}~\bibnamefont {Pritzl}}, \bibinfo
  {author} {\bibfnamefont {J.}~\bibnamefont {Hou}}, \bibinfo {author}
  {\bibfnamefont {V.}~\bibnamefont {Tileli}},\ and\ \bibinfo {author}
  {\bibfnamefont {H.~A.}\ \bibnamefont {Gasteiger}},\ }\bibfield  {title}
  {\bibinfo {title} {{Morphological Changes of Silicon Nanoparticles and the
  Influence of Cutoff Potentials in Silicon-Graphite Electrodes}},\ }\href
  {https://doi.org/10.1149/2.1261807jes} {\bibfield  {journal} {\bibinfo
  {journal} {Journal of The Electrochemical Society}\ }\textbf {\bibinfo
  {volume} {165}},\ \bibinfo {pages} {A1503} (\bibinfo {year}
  {2018})}\BibitemShut {NoStop}%
\bibitem [{\citenamefont {Kilchert}\ \emph {et~al.}(2024)\citenamefont
  {Kilchert}, \citenamefont {Schammer}, \citenamefont {Latz},\ and\
  \citenamefont {Horstmann}}]{Kilchert2024}%
  \BibitemOpen
  \bibfield  {author} {\bibinfo {author} {\bibfnamefont {F.}~\bibnamefont
  {Kilchert}}, \bibinfo {author} {\bibfnamefont {M.}~\bibnamefont {Schammer}},
  \bibinfo {author} {\bibfnamefont {A.}~\bibnamefont {Latz}},\ and\ \bibinfo
  {author} {\bibfnamefont {B.}~\bibnamefont {Horstmann}},\ }\bibfield  {title}
  {\bibinfo {title} {{Silicon Nanowires as Anodes for Lithium‐Ion Batteries:
  Full Cell Modeling}},\ }\bibfield  {journal} {\bibinfo  {journal} {Energy
  Technology}\ }\textbf {\bibinfo {volume} {2400206}},\ \href
  {https://doi.org/10.1002/ente.202400206} {10.1002/ente.202400206} (\bibinfo
  {year} {2024}),\ \Eprint {https://arxiv.org/abs/2401.16125} {2401.16125}
  \BibitemShut {NoStop}%
\bibitem [{\citenamefont {Pan}\ \emph {et~al.}(2019)\citenamefont {Pan},
  \citenamefont {Zou}, \citenamefont {Canova}, \citenamefont {Zhu},\ and\
  \citenamefont {Kim}}]{Pan2019}%
  \BibitemOpen
  \bibfield  {author} {\bibinfo {author} {\bibfnamefont {K.}~\bibnamefont
  {Pan}}, \bibinfo {author} {\bibfnamefont {F.}~\bibnamefont {Zou}}, \bibinfo
  {author} {\bibfnamefont {M.}~\bibnamefont {Canova}}, \bibinfo {author}
  {\bibfnamefont {Y.}~\bibnamefont {Zhu}},\ and\ \bibinfo {author}
  {\bibfnamefont {J.-H.}\ \bibnamefont {Kim}},\ }\bibfield  {title} {\bibinfo
  {title} {{Systematic electrochemical characterizations of Si and SiO anodes
  for high-capacity Li-Ion batteries}},\ }\href
  {https://doi.org/10.1016/j.jpowsour.2018.12.010} {\bibfield  {journal}
  {\bibinfo  {journal} {Journal of Power Sources}\ }\textbf {\bibinfo {volume}
  {413}},\ \bibinfo {pages} {20} (\bibinfo {year} {2019})}\BibitemShut
  {NoStop}%
\bibitem [{\citenamefont {Pan}(2020)}]{Pan2020}%
  \BibitemOpen
  \bibfield  {author} {\bibinfo {author} {\bibfnamefont {K.}~\bibnamefont
  {Pan}},\ }\emph {\bibinfo {title} {Dissertation}},\ \href@noop {} {Ph.D.
  thesis},\ \bibinfo  {school} {Ohio State University} (\bibinfo {year}
  {2020})\BibitemShut {NoStop}%
\bibitem [{\citenamefont {Bernard}\ \emph {et~al.}(2019)\citenamefont
  {Bernard}, \citenamefont {Alper}, \citenamefont {Haon}, \citenamefont
  {Herlin-Boime},\ and\ \citenamefont {Chandesris}}]{Bernard2019}%
  \BibitemOpen
  \bibfield  {author} {\bibinfo {author} {\bibfnamefont {P.}~\bibnamefont
  {Bernard}}, \bibinfo {author} {\bibfnamefont {J.~P.}\ \bibnamefont {Alper}},
  \bibinfo {author} {\bibfnamefont {C.}~\bibnamefont {Haon}}, \bibinfo {author}
  {\bibfnamefont {N.}~\bibnamefont {Herlin-Boime}},\ and\ \bibinfo {author}
  {\bibfnamefont {M.}~\bibnamefont {Chandesris}},\ }\bibfield  {title}
  {\bibinfo {title} {{Electrochemical analysis of silicon nanoparticle
  lithiation – Effect of crystallinity and carbon coating quantity}},\ }\href
  {https://doi.org/10.1016/j.jpowsour.2019.226769} {\bibfield  {journal}
  {\bibinfo  {journal} {Journal of Power Sources}\ }\textbf {\bibinfo {volume}
  {435}},\ \bibinfo {pages} {226769} (\bibinfo {year} {2019})}\BibitemShut
  {NoStop}%
\bibitem [{\citenamefont {Wycisk}\ \emph {et~al.}(2024)\citenamefont {Wycisk},
  \citenamefont {Mertin}, \citenamefont {Oldenburger}, \citenamefont {von
  Kessel},\ and\ \citenamefont {Latz}}]{Wycisk2023}%
  \BibitemOpen
  \bibfield  {author} {\bibinfo {author} {\bibfnamefont {D.}~\bibnamefont
  {Wycisk}}, \bibinfo {author} {\bibfnamefont {G.~K.}\ \bibnamefont {Mertin}},
  \bibinfo {author} {\bibfnamefont {M.}~\bibnamefont {Oldenburger}}, \bibinfo
  {author} {\bibfnamefont {O.}~\bibnamefont {von Kessel}},\ and\ \bibinfo
  {author} {\bibfnamefont {A.}~\bibnamefont {Latz}},\ }\bibfield  {title}
  {\bibinfo {title} {{Challenges of open-circuit voltage measurements for
  silicon-containing Li-Ion cells}},\ }\href
  {https://doi.org/10.1016/j.est.2024.111617} {\bibfield  {journal} {\bibinfo
  {journal} {Journal of Energy Storage}\ }\textbf {\bibinfo {volume} {89}},\
  \bibinfo {pages} {111617} (\bibinfo {year} {2024})}\BibitemShut {NoStop}%
\bibitem [{\citenamefont {McDowell}\ \emph
  {et~al.}(2013{\natexlab{a}})\citenamefont {McDowell}, \citenamefont {Lee},
  \citenamefont {Nix},\ and\ \citenamefont {Cui}}]{McDowell2013}%
  \BibitemOpen
  \bibfield  {author} {\bibinfo {author} {\bibfnamefont {M.~T.}\ \bibnamefont
  {McDowell}}, \bibinfo {author} {\bibfnamefont {S.~W.}\ \bibnamefont {Lee}},
  \bibinfo {author} {\bibfnamefont {W.~D.}\ \bibnamefont {Nix}},\ and\ \bibinfo
  {author} {\bibfnamefont {Y.}~\bibnamefont {Cui}},\ }\bibfield  {title}
  {\bibinfo {title} {{25th Anniversary Article: Understanding the Lithiation of
  Silicon and Other Alloying Anodes for Lithium‐Ion Batteries}},\ }\href
  {https://doi.org/10.1002/adma.201301795} {\bibfield  {journal} {\bibinfo
  {journal} {Advanced Materials}\ }\textbf {\bibinfo {volume} {25}},\ \bibinfo
  {pages} {4966} (\bibinfo {year} {2013}{\natexlab{a}})}\BibitemShut {NoStop}%
\bibitem [{\citenamefont {Wycisk}\ \emph {et~al.}(2023)\citenamefont {Wycisk},
  \citenamefont {Mertin}, \citenamefont {Oldenburger},\ and\ \citenamefont
  {Latz}}]{Wycisk2023Heat}%
  \BibitemOpen
  \bibfield  {author} {\bibinfo {author} {\bibfnamefont {D.}~\bibnamefont
  {Wycisk}}, \bibinfo {author} {\bibfnamefont {G.~K.}\ \bibnamefont {Mertin}},
  \bibinfo {author} {\bibfnamefont {M.}~\bibnamefont {Oldenburger}},\ and\
  \bibinfo {author} {\bibfnamefont {A.}~\bibnamefont {Latz}},\ }\bibfield
  {title} {\bibinfo {title} {{Analysis of heat generation due to open-circuit
  voltage hysteresis in lithium-ion cells}},\ }\href
  {https://doi.org/10.1016/j.est.2023.106817} {\bibfield  {journal} {\bibinfo
  {journal} {Journal of Energy Storage}\ }\textbf {\bibinfo {volume} {61}},\
  \bibinfo {pages} {106817} (\bibinfo {year} {2023})}\BibitemShut {NoStop}%
\bibitem [{\citenamefont {Sethuraman}\ \emph {et~al.}(2010)\citenamefont
  {Sethuraman}, \citenamefont {Chon}, \citenamefont {Shimshak}, \citenamefont
  {Srinivasan},\ and\ \citenamefont {Guduru}}]{Sethuraman2010a}%
  \BibitemOpen
  \bibfield  {author} {\bibinfo {author} {\bibfnamefont {V.~A.}\ \bibnamefont
  {Sethuraman}}, \bibinfo {author} {\bibfnamefont {M.~J.}\ \bibnamefont
  {Chon}}, \bibinfo {author} {\bibfnamefont {M.}~\bibnamefont {Shimshak}},
  \bibinfo {author} {\bibfnamefont {V.}~\bibnamefont {Srinivasan}},\ and\
  \bibinfo {author} {\bibfnamefont {P.~R.}\ \bibnamefont {Guduru}},\ }\bibfield
   {title} {\bibinfo {title} {{In situ measurements of stress evolution in
  silicon thin films during electrochemical lithiation and delithiation}},\
  }\href {https://doi.org/10.1016/j.jpowsour.2010.02.013} {\bibfield  {journal}
  {\bibinfo  {journal} {Journal of Power Sources}\ }\textbf {\bibinfo {volume}
  {195}},\ \bibinfo {pages} {5062} (\bibinfo {year} {2010})}\BibitemShut
  {NoStop}%
\bibitem [{\citenamefont {Lu}\ \emph {et~al.}(2016)\citenamefont {Lu},
  \citenamefont {Song}, \citenamefont {Zhang}, \citenamefont {Pan},
  \citenamefont {Cheng},\ and\ \citenamefont {Zhang}}]{Lu2016}%
  \BibitemOpen
  \bibfield  {author} {\bibinfo {author} {\bibfnamefont {B.}~\bibnamefont
  {Lu}}, \bibinfo {author} {\bibfnamefont {Y.}~\bibnamefont {Song}}, \bibinfo
  {author} {\bibfnamefont {Q.}~\bibnamefont {Zhang}}, \bibinfo {author}
  {\bibfnamefont {J.}~\bibnamefont {Pan}}, \bibinfo {author} {\bibfnamefont
  {Y.-T.}\ \bibnamefont {Cheng}},\ and\ \bibinfo {author} {\bibfnamefont
  {J.}~\bibnamefont {Zhang}},\ }\bibfield  {title} {\bibinfo {title} {{Voltage
  hysteresis of lithium ion batteries caused by mechanical stress}},\ }\href
  {https://doi.org/10.1039/C5CP06179B} {\bibfield  {journal} {\bibinfo
  {journal} {Physical Chemistry Chemical Physics}\ }\textbf {\bibinfo {volume}
  {18}},\ \bibinfo {pages} {4721} (\bibinfo {year} {2016})}\BibitemShut
  {NoStop}%
\bibitem [{\citenamefont {Chandrasekaran}\ \emph {et~al.}(2010)\citenamefont
  {Chandrasekaran}, \citenamefont {Magasinski}, \citenamefont {Yushin},\ and\
  \citenamefont {Fuller}}]{Chandrasekaran2010}%
  \BibitemOpen
  \bibfield  {author} {\bibinfo {author} {\bibfnamefont {R.}~\bibnamefont
  {Chandrasekaran}}, \bibinfo {author} {\bibfnamefont {A.}~\bibnamefont
  {Magasinski}}, \bibinfo {author} {\bibfnamefont {G.}~\bibnamefont {Yushin}},\
  and\ \bibinfo {author} {\bibfnamefont {T.~F.}\ \bibnamefont {Fuller}},\
  }\bibfield  {title} {\bibinfo {title} {{Analysis of Lithium
  Insertion/Deinsertion in a Silicon Electrode Particle at Room Temperature}},\
  }\href {https://doi.org/10.1149/1.3474225} {\bibfield  {journal} {\bibinfo
  {journal} {Journal of The Electrochemical Society}\ }\textbf {\bibinfo
  {volume} {157}},\ \bibinfo {pages} {A1139} (\bibinfo {year}
  {2010})}\BibitemShut {NoStop}%
\bibitem [{\citenamefont {Zhao}\ \emph {et~al.}(2011)\citenamefont {Zhao},
  \citenamefont {Pharr}, \citenamefont {Cai}, \citenamefont {Vlassak},\ and\
  \citenamefont {Suo}}]{Zhao2011}%
  \BibitemOpen
  \bibfield  {author} {\bibinfo {author} {\bibfnamefont {K.}~\bibnamefont
  {Zhao}}, \bibinfo {author} {\bibfnamefont {M.}~\bibnamefont {Pharr}},
  \bibinfo {author} {\bibfnamefont {S.}~\bibnamefont {Cai}}, \bibinfo {author}
  {\bibfnamefont {J.~J.}\ \bibnamefont {Vlassak}},\ and\ \bibinfo {author}
  {\bibfnamefont {Z.}~\bibnamefont {Suo}},\ }\bibfield  {title} {\bibinfo
  {title} {{Large Plastic Deformation in High‐Capacity Lithium‐Ion
  Batteries Caused by Charge and Discharge}},\ }\href
  {https://doi.org/10.1111/j.1551-2916.2011.04432.x} {\bibfield  {journal}
  {\bibinfo  {journal} {Journal of the American Ceramic Society}\ }\textbf
  {\bibinfo {volume} {94}},\ \bibinfo {pages} {s226} (\bibinfo {year}
  {2011})}\BibitemShut {NoStop}%
\bibitem [{\citenamefont {Cui}\ \emph {et~al.}(2012)\citenamefont {Cui},
  \citenamefont {Gao},\ and\ \citenamefont {Qu}}]{Cui2012}%
  \BibitemOpen
  \bibfield  {author} {\bibinfo {author} {\bibfnamefont {Z.}~\bibnamefont
  {Cui}}, \bibinfo {author} {\bibfnamefont {F.}~\bibnamefont {Gao}},\ and\
  \bibinfo {author} {\bibfnamefont {J.}~\bibnamefont {Qu}},\ }\bibfield
  {title} {\bibinfo {title} {{A finite deformation stress-dependent chemical
  potential and its applications to lithium ion batteries}},\ }\href
  {https://doi.org/10.1016/j.jmps.2012.03.008} {\bibfield  {journal} {\bibinfo
  {journal} {Journal of the Mechanics and Physics of Solids}\ }\textbf
  {\bibinfo {volume} {60}},\ \bibinfo {pages} {1280} (\bibinfo {year}
  {2012})}\BibitemShut {NoStop}%
\bibitem [{\citenamefont {McDowell}\ \emph
  {et~al.}(2013{\natexlab{b}})\citenamefont {McDowell}, \citenamefont {Lee},
  \citenamefont {Harris}, \citenamefont {Korgel}, \citenamefont {Wang},
  \citenamefont {Nix},\ and\ \citenamefont {Cui}}]{McDowell2013Trafo}%
  \BibitemOpen
  \bibfield  {author} {\bibinfo {author} {\bibfnamefont {M.~T.}\ \bibnamefont
  {McDowell}}, \bibinfo {author} {\bibfnamefont {S.~W.}\ \bibnamefont {Lee}},
  \bibinfo {author} {\bibfnamefont {J.~T.}\ \bibnamefont {Harris}}, \bibinfo
  {author} {\bibfnamefont {B.~A.}\ \bibnamefont {Korgel}}, \bibinfo {author}
  {\bibfnamefont {C.}~\bibnamefont {Wang}}, \bibinfo {author} {\bibfnamefont
  {W.~D.}\ \bibnamefont {Nix}},\ and\ \bibinfo {author} {\bibfnamefont
  {Y.}~\bibnamefont {Cui}},\ }\bibfield  {title} {\bibinfo {title} {{In Situ
  TEM of Two-Phase Lithiation of Amorphous Silicon Nanospheres}},\ }\href
  {https://doi.org/10.1021/nl3044508} {\bibfield  {journal} {\bibinfo
  {journal} {Nano Letters}\ }\textbf {\bibinfo {volume} {13}},\ \bibinfo
  {pages} {758} (\bibinfo {year} {2013}{\natexlab{b}})}\BibitemShut {NoStop}%
\bibitem [{\citenamefont {Sethuraman}\ \emph {et~al.}(2013)\citenamefont
  {Sethuraman}, \citenamefont {Srinivasan},\ and\ \citenamefont
  {Newman}}]{Sethuraman2013}%
  \BibitemOpen
  \bibfield  {author} {\bibinfo {author} {\bibfnamefont {V.~A.}\ \bibnamefont
  {Sethuraman}}, \bibinfo {author} {\bibfnamefont {V.}~\bibnamefont
  {Srinivasan}},\ and\ \bibinfo {author} {\bibfnamefont {J.}~\bibnamefont
  {Newman}},\ }\bibfield  {title} {\bibinfo {title} {{Analysis of
  Electrochemical Lithiation and Delithiation Kinetics in Silicon}},\ }\href
  {https://doi.org/10.1149/2.008303jes} {\bibfield  {journal} {\bibinfo
  {journal} {Journal of The Electrochemical Society}\ }\textbf {\bibinfo
  {volume} {160}},\ \bibinfo {pages} {A394} (\bibinfo {year}
  {2013})}\BibitemShut {NoStop}%
\bibitem [{\citenamefont {K{\"{o}}bbing}\ \emph {et~al.}(2024)\citenamefont
  {K{\"{o}}bbing}, \citenamefont {Latz},\ and\ \citenamefont
  {Horstmann}}]{Koebbing2023Voltage}%
  \BibitemOpen
  \bibfield  {author} {\bibinfo {author} {\bibfnamefont {L.}~\bibnamefont
  {K{\"{o}}bbing}}, \bibinfo {author} {\bibfnamefont {A.}~\bibnamefont
  {Latz}},\ and\ \bibinfo {author} {\bibfnamefont {B.}~\bibnamefont
  {Horstmann}},\ }\bibfield  {title} {\bibinfo {title} {{Voltage Hysteresis of
  Silicon Nanoparticles: Chemo‐Mechanical Particle‐SEI Model}},\ }\href
  {https://doi.org/10.1002/adfm.202308818} {\bibfield  {journal} {\bibinfo
  {journal} {Advanced Functional Materials}\ }\textbf {\bibinfo {volume}
  {34}},\ \bibinfo {pages} {2308818} (\bibinfo {year} {2024})}\BibitemShut
  {NoStop}%
\bibitem [{\citenamefont {Horstmann}\ \emph {et~al.}(2019)\citenamefont
  {Horstmann}, \citenamefont {Single},\ and\ \citenamefont
  {Latz}}]{Horstmann2019}%
  \BibitemOpen
  \bibfield  {author} {\bibinfo {author} {\bibfnamefont {B.}~\bibnamefont
  {Horstmann}}, \bibinfo {author} {\bibfnamefont {F.}~\bibnamefont {Single}},\
  and\ \bibinfo {author} {\bibfnamefont {A.}~\bibnamefont {Latz}},\ }\bibfield
  {title} {\bibinfo {title} {{Review on multi-scale models of solid-electrolyte
  interphase formation}},\ }\href
  {https://doi.org/10.1016/j.coelec.2018.10.013} {\bibfield  {journal}
  {\bibinfo  {journal} {Current Opinion in Electrochemistry}\ }\textbf
  {\bibinfo {volume} {13}},\ \bibinfo {pages} {61} (\bibinfo {year}
  {2019})}\BibitemShut {NoStop}%
\bibitem [{\citenamefont {Zhang}\ \emph {et~al.}(2021)\citenamefont {Zhang},
  \citenamefont {Weng}, \citenamefont {Yang}, \citenamefont {Li}, \citenamefont
  {Li}, \citenamefont {Su}, \citenamefont {Gu}, \citenamefont {Wang},
  \citenamefont {Wang},\ and\ \citenamefont {Chen}}]{Zhang2021interplay}%
  \BibitemOpen
  \bibfield  {author} {\bibinfo {author} {\bibfnamefont {X.}~\bibnamefont
  {Zhang}}, \bibinfo {author} {\bibfnamefont {S.}~\bibnamefont {Weng}},
  \bibinfo {author} {\bibfnamefont {G.}~\bibnamefont {Yang}}, \bibinfo {author}
  {\bibfnamefont {Y.}~\bibnamefont {Li}}, \bibinfo {author} {\bibfnamefont
  {H.}~\bibnamefont {Li}}, \bibinfo {author} {\bibfnamefont {D.}~\bibnamefont
  {Su}}, \bibinfo {author} {\bibfnamefont {L.}~\bibnamefont {Gu}}, \bibinfo
  {author} {\bibfnamefont {Z.}~\bibnamefont {Wang}}, \bibinfo {author}
  {\bibfnamefont {X.}~\bibnamefont {Wang}},\ and\ \bibinfo {author}
  {\bibfnamefont {L.}~\bibnamefont {Chen}},\ }\bibfield  {title} {\bibinfo
  {title} {{Interplay between solid-electrolyte interphase and (in)active LixSi
  in silicon anode}},\ }\href {https://doi.org/10.1016/j.xcrp.2021.100668}
  {\bibfield  {journal} {\bibinfo  {journal} {Cell Reports Physical Science}\
  }\textbf {\bibinfo {volume} {2}},\ \bibinfo {pages} {100668} (\bibinfo {year}
  {2021})}\BibitemShut {NoStop}%
\bibitem [{\citenamefont {Nie}\ \emph {et~al.}(2013)\citenamefont {Nie},
  \citenamefont {Abraham}, \citenamefont {Chen}, \citenamefont {Bose},\ and\
  \citenamefont {Lucht}}]{Nie2013}%
  \BibitemOpen
  \bibfield  {author} {\bibinfo {author} {\bibfnamefont {M.}~\bibnamefont
  {Nie}}, \bibinfo {author} {\bibfnamefont {D.~P.}\ \bibnamefont {Abraham}},
  \bibinfo {author} {\bibfnamefont {Y.}~\bibnamefont {Chen}}, \bibinfo {author}
  {\bibfnamefont {A.}~\bibnamefont {Bose}},\ and\ \bibinfo {author}
  {\bibfnamefont {B.~L.}\ \bibnamefont {Lucht}},\ }\bibfield  {title} {\bibinfo
  {title} {{Silicon Solid Electrolyte Interphase (SEI) of Lithium Ion Battery
  Characterized by Microscopy and Spectroscopy}},\ }\href
  {https://doi.org/10.1021/jp404155y} {\bibfield  {journal} {\bibinfo
  {journal} {The Journal of Physical Chemistry C}\ }\textbf {\bibinfo {volume}
  {117}},\ \bibinfo {pages} {13403} (\bibinfo {year} {2013})}\BibitemShut
  {NoStop}%
\bibitem [{\citenamefont {Verma}\ \emph {et~al.}(2010)\citenamefont {Verma},
  \citenamefont {Maire},\ and\ \citenamefont {Nov{\'{a}}k}}]{Verma2010}%
  \BibitemOpen
  \bibfield  {author} {\bibinfo {author} {\bibfnamefont {P.}~\bibnamefont
  {Verma}}, \bibinfo {author} {\bibfnamefont {P.}~\bibnamefont {Maire}},\ and\
  \bibinfo {author} {\bibfnamefont {P.}~\bibnamefont {Nov{\'{a}}k}},\
  }\bibfield  {title} {\bibinfo {title} {{A review of the features and analyses
  of the solid electrolyte interphase in Li-ion batteries}},\ }\href
  {https://doi.org/10.1016/j.electacta.2010.05.072} {\bibfield  {journal}
  {\bibinfo  {journal} {Electrochimica Acta}\ }\textbf {\bibinfo {volume}
  {55}},\ \bibinfo {pages} {6332} (\bibinfo {year} {2010})}\BibitemShut
  {NoStop}%
\bibitem [{\citenamefont {Peled}\ and\ \citenamefont
  {Menkin}(2017)}]{Peled2017}%
  \BibitemOpen
  \bibfield  {author} {\bibinfo {author} {\bibfnamefont {E.}~\bibnamefont
  {Peled}}\ and\ \bibinfo {author} {\bibfnamefont {S.}~\bibnamefont {Menkin}},\
  }\bibfield  {title} {\bibinfo {title} {{Review—SEI: Past, Present and
  Future}},\ }\href {https://doi.org/10.1149/2.1441707jes} {\bibfield
  {journal} {\bibinfo  {journal} {Journal of The Electrochemical Society}\
  }\textbf {\bibinfo {volume} {164}},\ \bibinfo {pages} {A1703} (\bibinfo
  {year} {2017})}\BibitemShut {NoStop}%
\bibitem [{\citenamefont {K{\"{o}}bbing}\ \emph {et~al.}(2023)\citenamefont
  {K{\"{o}}bbing}, \citenamefont {Latz},\ and\ \citenamefont
  {Horstmann}}]{Koebbing2023Growth}%
  \BibitemOpen
  \bibfield  {author} {\bibinfo {author} {\bibfnamefont {L.}~\bibnamefont
  {K{\"{o}}bbing}}, \bibinfo {author} {\bibfnamefont {A.}~\bibnamefont
  {Latz}},\ and\ \bibinfo {author} {\bibfnamefont {B.}~\bibnamefont
  {Horstmann}},\ }\bibfield  {title} {\bibinfo {title} {{Growth of the
  solid-electrolyte interphase: Electron diffusion versus solvent diffusion}},\
  }\href {https://doi.org/10.1016/j.jpowsour.2023.232651} {\bibfield  {journal}
  {\bibinfo  {journal} {Journal of Power Sources}\ }\textbf {\bibinfo {volume}
  {561}},\ \bibinfo {pages} {232651} (\bibinfo {year} {2023})}\BibitemShut
  {NoStop}%
\bibitem [{\citenamefont {von Kolzenberg}\ \emph {et~al.}(2020)\citenamefont
  {von Kolzenberg}, \citenamefont {Latz},\ and\ \citenamefont
  {Horstmann}}]{Kolzenberg2020}%
  \BibitemOpen
  \bibfield  {author} {\bibinfo {author} {\bibfnamefont {L.}~\bibnamefont {von
  Kolzenberg}}, \bibinfo {author} {\bibfnamefont {A.}~\bibnamefont {Latz}},\
  and\ \bibinfo {author} {\bibfnamefont {B.}~\bibnamefont {Horstmann}},\
  }\bibfield  {title} {\bibinfo {title} {{Solid–Electrolyte Interphase During
  Battery Cycling: Theory of Growth Regimes}},\ }\href
  {https://doi.org/10.1002/cssc.202000867} {\bibfield  {journal} {\bibinfo
  {journal} {ChemSusChem}\ }\textbf {\bibinfo {volume} {13}},\ \bibinfo {pages}
  {3901} (\bibinfo {year} {2020})}\BibitemShut {NoStop}%
\bibitem [{\citenamefont {von Kolzenberg}\ \emph {et~al.}(2022)\citenamefont
  {von Kolzenberg}, \citenamefont {Latz},\ and\ \citenamefont
  {Horstmann}}]{Kolzenberg2021}%
  \BibitemOpen
  \bibfield  {author} {\bibinfo {author} {\bibfnamefont {L.}~\bibnamefont {von
  Kolzenberg}}, \bibinfo {author} {\bibfnamefont {A.}~\bibnamefont {Latz}},\
  and\ \bibinfo {author} {\bibfnamefont {B.}~\bibnamefont {Horstmann}},\
  }\bibfield  {title} {\bibinfo {title} {{Chemo‐Mechanical Model of SEI
  Growth on Silicon Electrode Particles}},\ }\href
  {https://doi.org/10.1002/batt.202100216} {\bibfield  {journal} {\bibinfo
  {journal} {Batteries \& Supercaps}\ }\textbf {\bibinfo {volume} {5}},\
  \bibinfo {pages} {1} (\bibinfo {year} {2022})}\BibitemShut {NoStop}%
\bibitem [{\citenamefont {Guo}\ \emph {et~al.}(2020)\citenamefont {Guo},
  \citenamefont {Kumar}, \citenamefont {Xiao}, \citenamefont {Sheldon},\ and\
  \citenamefont {Gao}}]{Guo2020}%
  \BibitemOpen
  \bibfield  {author} {\bibinfo {author} {\bibfnamefont {K.}~\bibnamefont
  {Guo}}, \bibinfo {author} {\bibfnamefont {R.}~\bibnamefont {Kumar}}, \bibinfo
  {author} {\bibfnamefont {X.}~\bibnamefont {Xiao}}, \bibinfo {author}
  {\bibfnamefont {B.~W.}\ \bibnamefont {Sheldon}},\ and\ \bibinfo {author}
  {\bibfnamefont {H.}~\bibnamefont {Gao}},\ }\bibfield  {title} {\bibinfo
  {title} {{Failure progression in the solid electrolyte interphase (SEI) on
  silicon electrodes}},\ }\href {https://doi.org/10.1016/j.nanoen.2019.104257}
  {\bibfield  {journal} {\bibinfo  {journal} {Nano Energy}\ }\textbf {\bibinfo
  {volume} {68}},\ \bibinfo {pages} {104257} (\bibinfo {year}
  {2020})}\BibitemShut {NoStop}%
\bibitem [{\citenamefont {Lee}\ \emph {et~al.}(2024)\citenamefont {Lee},
  \citenamefont {Jeong}, \citenamefont {Ha}, \citenamefont {Kim},\ and\
  \citenamefont {Choi}}]{Lee2024}%
  \BibitemOpen
  \bibfield  {author} {\bibinfo {author} {\bibfnamefont {J.}~\bibnamefont
  {Lee}}, \bibinfo {author} {\bibfnamefont {J.-y.}\ \bibnamefont {Jeong}},
  \bibinfo {author} {\bibfnamefont {J.}~\bibnamefont {Ha}}, \bibinfo {author}
  {\bibfnamefont {Y.-t.}\ \bibnamefont {Kim}},\ and\ \bibinfo {author}
  {\bibfnamefont {J.}~\bibnamefont {Choi}},\ }\bibfield  {title} {\bibinfo
  {title} {{Understanding solid electrolyte interface formation on graphite and
  silicon anodes in lithium-ion batteries: Exploring the role of fluoroethylene
  carbonate}},\ }\href {https://doi.org/10.1016/j.elecom.2024.107708}
  {\bibfield  {journal} {\bibinfo  {journal} {Electrochemistry Communications}\
  }\textbf {\bibinfo {volume} {163}},\ \bibinfo {pages} {107708} (\bibinfo
  {year} {2024})}\BibitemShut {NoStop}%
\bibitem [{\citenamefont {Cao}\ \emph {et~al.}(2019)\citenamefont {Cao},
  \citenamefont {Abate}, \citenamefont {Sivonxay}, \citenamefont {Shyam},
  \citenamefont {Jia}, \citenamefont {Moritz}, \citenamefont {Devereaux},
  \citenamefont {Persson}, \citenamefont {Steinr{\"{u}}ck},\ and\ \citenamefont
  {Toney}}]{Cao2019}%
  \BibitemOpen
  \bibfield  {author} {\bibinfo {author} {\bibfnamefont {C.}~\bibnamefont
  {Cao}}, \bibinfo {author} {\bibfnamefont {I.~I.}\ \bibnamefont {Abate}},
  \bibinfo {author} {\bibfnamefont {E.}~\bibnamefont {Sivonxay}}, \bibinfo
  {author} {\bibfnamefont {B.}~\bibnamefont {Shyam}}, \bibinfo {author}
  {\bibfnamefont {C.}~\bibnamefont {Jia}}, \bibinfo {author} {\bibfnamefont
  {B.}~\bibnamefont {Moritz}}, \bibinfo {author} {\bibfnamefont {T.~P.}\
  \bibnamefont {Devereaux}}, \bibinfo {author} {\bibfnamefont {K.~A.}\
  \bibnamefont {Persson}}, \bibinfo {author} {\bibfnamefont {H.-G.}\
  \bibnamefont {Steinr{\"{u}}ck}},\ and\ \bibinfo {author} {\bibfnamefont
  {M.~F.}\ \bibnamefont {Toney}},\ }\bibfield  {title} {\bibinfo {title}
  {{Solid Electrolyte Interphase on Native Oxide-Terminated Silicon Anodes for
  Li-Ion Batteries}},\ }\href {https://doi.org/10.1016/j.joule.2018.12.013}
  {\bibfield  {journal} {\bibinfo  {journal} {Joule}\ }\textbf {\bibinfo
  {volume} {3}},\ \bibinfo {pages} {762} (\bibinfo {year} {2019})}\BibitemShut
  {NoStop}%
\bibitem [{\citenamefont {Li}\ \emph {et~al.}(2019)\citenamefont {Li},
  \citenamefont {Liu}, \citenamefont {Han}, \citenamefont {Xiang},
  \citenamefont {Zhong}, \citenamefont {Wang}, \citenamefont {Zheng},
  \citenamefont {Zhou},\ and\ \citenamefont {Yang}}]{Li2019}%
  \BibitemOpen
  \bibfield  {author} {\bibinfo {author} {\bibfnamefont {Q.}~\bibnamefont
  {Li}}, \bibinfo {author} {\bibfnamefont {X.}~\bibnamefont {Liu}}, \bibinfo
  {author} {\bibfnamefont {X.}~\bibnamefont {Han}}, \bibinfo {author}
  {\bibfnamefont {Y.}~\bibnamefont {Xiang}}, \bibinfo {author} {\bibfnamefont
  {G.}~\bibnamefont {Zhong}}, \bibinfo {author} {\bibfnamefont
  {J.}~\bibnamefont {Wang}}, \bibinfo {author} {\bibfnamefont {B.}~\bibnamefont
  {Zheng}}, \bibinfo {author} {\bibfnamefont {J.}~\bibnamefont {Zhou}},\ and\
  \bibinfo {author} {\bibfnamefont {Y.}~\bibnamefont {Yang}},\ }\bibfield
  {title} {\bibinfo {title} {{Identification of the Solid Electrolyte Interface
  on the Si/C Composite Anode with FEC as the Additive}},\ }\href
  {https://doi.org/10.1021/acsami.8b22221} {\bibfield  {journal} {\bibinfo
  {journal} {ACS Applied Materials \& Interfaces}\ }\textbf {\bibinfo {volume}
  {11}},\ \bibinfo {pages} {14066} (\bibinfo {year} {2019})}\BibitemShut
  {NoStop}%
\bibitem [{\citenamefont {Chen}\ \emph {et~al.}(2020)\citenamefont {Chen},
  \citenamefont {Fan}, \citenamefont {Li}, \citenamefont {Yang}, \citenamefont
  {Khoshi}, \citenamefont {Xu}, \citenamefont {Hwang}, \citenamefont {Chen},
  \citenamefont {Ji}, \citenamefont {Yang}, \citenamefont {He}, \citenamefont
  {Wang}, \citenamefont {Garfunkel}, \citenamefont {Su}, \citenamefont
  {Borodin},\ and\ \citenamefont {Wang}}]{Chen2020}%
  \BibitemOpen
  \bibfield  {author} {\bibinfo {author} {\bibfnamefont {J.}~\bibnamefont
  {Chen}}, \bibinfo {author} {\bibfnamefont {X.}~\bibnamefont {Fan}}, \bibinfo
  {author} {\bibfnamefont {Q.}~\bibnamefont {Li}}, \bibinfo {author}
  {\bibfnamefont {H.}~\bibnamefont {Yang}}, \bibinfo {author} {\bibfnamefont
  {M.~R.}\ \bibnamefont {Khoshi}}, \bibinfo {author} {\bibfnamefont
  {Y.}~\bibnamefont {Xu}}, \bibinfo {author} {\bibfnamefont {S.}~\bibnamefont
  {Hwang}}, \bibinfo {author} {\bibfnamefont {L.}~\bibnamefont {Chen}},
  \bibinfo {author} {\bibfnamefont {X.}~\bibnamefont {Ji}}, \bibinfo {author}
  {\bibfnamefont {C.}~\bibnamefont {Yang}}, \bibinfo {author} {\bibfnamefont
  {H.}~\bibnamefont {He}}, \bibinfo {author} {\bibfnamefont {C.}~\bibnamefont
  {Wang}}, \bibinfo {author} {\bibfnamefont {E.}~\bibnamefont {Garfunkel}},
  \bibinfo {author} {\bibfnamefont {D.}~\bibnamefont {Su}}, \bibinfo {author}
  {\bibfnamefont {O.}~\bibnamefont {Borodin}},\ and\ \bibinfo {author}
  {\bibfnamefont {C.}~\bibnamefont {Wang}},\ }\bibfield  {title} {\bibinfo
  {title} {{Electrolyte design for LiF-rich solid–electrolyte interfaces to
  enable high-performance microsized alloy anodes for batteries}},\ }\href
  {https://doi.org/10.1038/s41560-020-0601-1} {\bibfield  {journal} {\bibinfo
  {journal} {Nature Energy}\ }\textbf {\bibinfo {volume} {5}},\ \bibinfo
  {pages} {386} (\bibinfo {year} {2020})}\BibitemShut {NoStop}%
\bibitem [{\citenamefont {He}\ \emph {et~al.}(2021)\citenamefont {He},
  \citenamefont {Jiang}, \citenamefont {Chen}, \citenamefont {Xu},
  \citenamefont {Jia}, \citenamefont {Yi}, \citenamefont {Xue}, \citenamefont
  {Song}, \citenamefont {Genc}, \citenamefont {Bouchet-Marquis}, \citenamefont
  {Pullan}, \citenamefont {Tessner}, \citenamefont {Yoo}, \citenamefont {Li},
  \citenamefont {Zhang}, \citenamefont {Zhang},\ and\ \citenamefont
  {Wang}}]{He2021}%
  \BibitemOpen
  \bibfield  {author} {\bibinfo {author} {\bibfnamefont {Y.}~\bibnamefont
  {He}}, \bibinfo {author} {\bibfnamefont {L.}~\bibnamefont {Jiang}}, \bibinfo
  {author} {\bibfnamefont {T.}~\bibnamefont {Chen}}, \bibinfo {author}
  {\bibfnamefont {Y.}~\bibnamefont {Xu}}, \bibinfo {author} {\bibfnamefont
  {H.}~\bibnamefont {Jia}}, \bibinfo {author} {\bibfnamefont {R.}~\bibnamefont
  {Yi}}, \bibinfo {author} {\bibfnamefont {D.}~\bibnamefont {Xue}}, \bibinfo
  {author} {\bibfnamefont {M.}~\bibnamefont {Song}}, \bibinfo {author}
  {\bibfnamefont {A.}~\bibnamefont {Genc}}, \bibinfo {author} {\bibfnamefont
  {C.}~\bibnamefont {Bouchet-Marquis}}, \bibinfo {author} {\bibfnamefont
  {L.}~\bibnamefont {Pullan}}, \bibinfo {author} {\bibfnamefont
  {T.}~\bibnamefont {Tessner}}, \bibinfo {author} {\bibfnamefont
  {J.}~\bibnamefont {Yoo}}, \bibinfo {author} {\bibfnamefont {X.}~\bibnamefont
  {Li}}, \bibinfo {author} {\bibfnamefont {J.-G.}\ \bibnamefont {Zhang}},
  \bibinfo {author} {\bibfnamefont {S.}~\bibnamefont {Zhang}},\ and\ \bibinfo
  {author} {\bibfnamefont {C.}~\bibnamefont {Wang}},\ }\bibfield  {title}
  {\bibinfo {title} {{Progressive growth of the solid–electrolyte interphase
  towards the Si anode interior causes capacity fading}},\ }\href
  {https://doi.org/10.1038/s41565-021-00947-8} {\bibfield  {journal} {\bibinfo
  {journal} {Nature Nanotechnology}\ }\textbf {\bibinfo {volume} {16}},\
  \bibinfo {pages} {1113} (\bibinfo {year} {2021})}\BibitemShut {NoStop}%
\bibitem [{\citenamefont {Schnabel}\ \emph {et~al.}(2020)\citenamefont
  {Schnabel}, \citenamefont {Harvey}, \citenamefont {Arca}, \citenamefont
  {Stetson}, \citenamefont {Teeter}, \citenamefont {Ban},\ and\ \citenamefont
  {Stradins}}]{Schnabel2020}%
  \BibitemOpen
  \bibfield  {author} {\bibinfo {author} {\bibfnamefont {M.}~\bibnamefont
  {Schnabel}}, \bibinfo {author} {\bibfnamefont {S.~P.}\ \bibnamefont
  {Harvey}}, \bibinfo {author} {\bibfnamefont {E.}~\bibnamefont {Arca}},
  \bibinfo {author} {\bibfnamefont {C.}~\bibnamefont {Stetson}}, \bibinfo
  {author} {\bibfnamefont {G.}~\bibnamefont {Teeter}}, \bibinfo {author}
  {\bibfnamefont {C.}~\bibnamefont {Ban}},\ and\ \bibinfo {author}
  {\bibfnamefont {P.}~\bibnamefont {Stradins}},\ }\bibfield  {title} {\bibinfo
  {title} {{Surface SiO 2 Thickness Controls Uniform-to-Localized Transition in
  Lithiation of Silicon Anodes for Lithium-Ion Batteries}},\ }\href
  {https://doi.org/10.1021/acsami.0c03158} {\bibfield  {journal} {\bibinfo
  {journal} {ACS Applied Materials \& Interfaces}\ }\textbf {\bibinfo {volume}
  {12}},\ \bibinfo {pages} {27017} (\bibinfo {year} {2020})}\BibitemShut
  {NoStop}%
\bibitem [{\citenamefont {Schroder}\ \emph {et~al.}(2014)\citenamefont
  {Schroder}, \citenamefont {Dylla}, \citenamefont {Harris}, \citenamefont
  {Webb},\ and\ \citenamefont {Stevenson}}]{Schroder2014}%
  \BibitemOpen
  \bibfield  {author} {\bibinfo {author} {\bibfnamefont {K.~W.}\ \bibnamefont
  {Schroder}}, \bibinfo {author} {\bibfnamefont {A.~G.}\ \bibnamefont {Dylla}},
  \bibinfo {author} {\bibfnamefont {S.~J.}\ \bibnamefont {Harris}}, \bibinfo
  {author} {\bibfnamefont {L.~J.}\ \bibnamefont {Webb}},\ and\ \bibinfo
  {author} {\bibfnamefont {K.~J.}\ \bibnamefont {Stevenson}},\ }\bibfield
  {title} {\bibinfo {title} {{Role of Surface Oxides in the Formation of
  Solid–Electrolyte Interphases at Silicon Electrodes for Lithium-Ion
  Batteries}},\ }\href {https://doi.org/10.1021/am506517j} {\bibfield
  {journal} {\bibinfo  {journal} {ACS Applied Materials \& Interfaces}\
  }\textbf {\bibinfo {volume} {6}},\ \bibinfo {pages} {21510} (\bibinfo {year}
  {2014})}\BibitemShut {NoStop}%
\bibitem [{\citenamefont {Deringer}\ \emph {et~al.}(2021)\citenamefont
  {Deringer}, \citenamefont {Bernstein}, \citenamefont {Cs{\'{a}}nyi},
  \citenamefont {{Ben Mahmoud}}, \citenamefont {Ceriotti}, \citenamefont
  {Wilson}, \citenamefont {Drabold},\ and\ \citenamefont
  {Elliott}}]{Deringer2021}%
  \BibitemOpen
  \bibfield  {author} {\bibinfo {author} {\bibfnamefont {V.~L.}\ \bibnamefont
  {Deringer}}, \bibinfo {author} {\bibfnamefont {N.}~\bibnamefont {Bernstein}},
  \bibinfo {author} {\bibfnamefont {G.}~\bibnamefont {Cs{\'{a}}nyi}}, \bibinfo
  {author} {\bibfnamefont {C.}~\bibnamefont {{Ben Mahmoud}}}, \bibinfo {author}
  {\bibfnamefont {M.}~\bibnamefont {Ceriotti}}, \bibinfo {author}
  {\bibfnamefont {M.}~\bibnamefont {Wilson}}, \bibinfo {author} {\bibfnamefont
  {D.~A.}\ \bibnamefont {Drabold}},\ and\ \bibinfo {author} {\bibfnamefont
  {S.~R.}\ \bibnamefont {Elliott}},\ }\bibfield  {title} {\bibinfo {title}
  {{Origins of structural and electronic transitions in disordered silicon}},\
  }\href {https://doi.org/10.1038/s41586-020-03072-z} {\bibfield  {journal}
  {\bibinfo  {journal} {Nature}\ }\textbf {\bibinfo {volume} {589}},\ \bibinfo
  {pages} {59} (\bibinfo {year} {2021})}\BibitemShut {NoStop}%
\bibitem [{\citenamefont {Chevrier}\ \emph {et~al.}(2014)\citenamefont
  {Chevrier}, \citenamefont {Liu}, \citenamefont {Le}, \citenamefont {Lund},
  \citenamefont {Molla}, \citenamefont {Reimer}, \citenamefont {Krause},
  \citenamefont {Jensen}, \citenamefont {Figgemeier},\ and\ \citenamefont
  {Eberman}}]{Chevrier2014}%
  \BibitemOpen
  \bibfield  {author} {\bibinfo {author} {\bibfnamefont {V.~L.}\ \bibnamefont
  {Chevrier}}, \bibinfo {author} {\bibfnamefont {L.}~\bibnamefont {Liu}},
  \bibinfo {author} {\bibfnamefont {D.~B.}\ \bibnamefont {Le}}, \bibinfo
  {author} {\bibfnamefont {J.}~\bibnamefont {Lund}}, \bibinfo {author}
  {\bibfnamefont {B.}~\bibnamefont {Molla}}, \bibinfo {author} {\bibfnamefont
  {K.}~\bibnamefont {Reimer}}, \bibinfo {author} {\bibfnamefont {L.~J.}\
  \bibnamefont {Krause}}, \bibinfo {author} {\bibfnamefont {L.~D.}\
  \bibnamefont {Jensen}}, \bibinfo {author} {\bibfnamefont {E.}~\bibnamefont
  {Figgemeier}},\ and\ \bibinfo {author} {\bibfnamefont {K.~W.}\ \bibnamefont
  {Eberman}},\ }\bibfield  {title} {\bibinfo {title} {{Evaluating Si-Based
  Materials for Li-Ion Batteries in Commercially Relevant Negative
  Electrodes}},\ }\href {https://doi.org/10.1149/2.066405jes} {\bibfield
  {journal} {\bibinfo  {journal} {Journal of The Electrochemical Society}\
  }\textbf {\bibinfo {volume} {161}},\ \bibinfo {pages} {A783} (\bibinfo {year}
  {2014})}\BibitemShut {NoStop}%
\bibitem [{\citenamefont {Kitada}\ \emph {et~al.}(2019)\citenamefont {Kitada},
  \citenamefont {Pecher}, \citenamefont {Magusin}, \citenamefont {Groh},
  \citenamefont {Weatherup},\ and\ \citenamefont {Grey}}]{Kitada2019}%
  \BibitemOpen
  \bibfield  {author} {\bibinfo {author} {\bibfnamefont {K.}~\bibnamefont
  {Kitada}}, \bibinfo {author} {\bibfnamefont {O.}~\bibnamefont {Pecher}},
  \bibinfo {author} {\bibfnamefont {P.~C. M.~M.}\ \bibnamefont {Magusin}},
  \bibinfo {author} {\bibfnamefont {M.~F.}\ \bibnamefont {Groh}}, \bibinfo
  {author} {\bibfnamefont {R.~S.}\ \bibnamefont {Weatherup}},\ and\ \bibinfo
  {author} {\bibfnamefont {C.~P.}\ \bibnamefont {Grey}},\ }\bibfield  {title}
  {\bibinfo {title} {{Unraveling the Reaction Mechanisms of SiO Anodes for
  Li-Ion Batteries by Combining in Situ 7 Li and ex Situ 7 Li/ 29 Si
  Solid-State NMR Spectroscopy}},\ }\href
  {https://doi.org/10.1021/jacs.9b01589} {\bibfield  {journal} {\bibinfo
  {journal} {Journal of the American Chemical Society}\ }\textbf {\bibinfo
  {volume} {141}},\ \bibinfo {pages} {7014} (\bibinfo {year}
  {2019})}\BibitemShut {NoStop}%
\bibitem [{\citenamefont {Wang}\ \emph {et~al.}(2020)\citenamefont {Wang},
  \citenamefont {Wang}, \citenamefont {Liu}, \citenamefont {Lu}, \citenamefont
  {Chu}, \citenamefont {Liu}, \citenamefont {Guo}, \citenamefont {Yu},
  \citenamefont {Luo}, \citenamefont {Ren}, \citenamefont {Chen},\ and\
  \citenamefont {Li}}]{Wang2020}%
  \BibitemOpen
  \bibfield  {author} {\bibinfo {author} {\bibfnamefont {J.}~\bibnamefont
  {Wang}}, \bibinfo {author} {\bibfnamefont {X.}~\bibnamefont {Wang}}, \bibinfo
  {author} {\bibfnamefont {B.}~\bibnamefont {Liu}}, \bibinfo {author}
  {\bibfnamefont {H.}~\bibnamefont {Lu}}, \bibinfo {author} {\bibfnamefont
  {G.}~\bibnamefont {Chu}}, \bibinfo {author} {\bibfnamefont {J.}~\bibnamefont
  {Liu}}, \bibinfo {author} {\bibfnamefont {Y.-G.}\ \bibnamefont {Guo}},
  \bibinfo {author} {\bibfnamefont {X.}~\bibnamefont {Yu}}, \bibinfo {author}
  {\bibfnamefont {F.}~\bibnamefont {Luo}}, \bibinfo {author} {\bibfnamefont
  {Y.}~\bibnamefont {Ren}}, \bibinfo {author} {\bibfnamefont {L.}~\bibnamefont
  {Chen}},\ and\ \bibinfo {author} {\bibfnamefont {H.}~\bibnamefont {Li}},\
  }\bibfield  {title} {\bibinfo {title} {{Size effect on the growth and
  pulverization behavior of Si nanodomains in SiO anode}},\ }\href
  {https://doi.org/10.1016/j.nanoen.2020.105101} {\bibfield  {journal}
  {\bibinfo  {journal} {Nano Energy}\ }\textbf {\bibinfo {volume} {78}},\
  \bibinfo {pages} {105101} (\bibinfo {year} {2020})}\BibitemShut {NoStop}%
\bibitem [{\citenamefont {Garofalo}\ \emph {et~al.}(1963)\citenamefont
  {Garofalo}, \citenamefont {Richmond}, \citenamefont {Domis},\ and\
  \citenamefont {von Gemmingen}}]{Garofalo1963}%
  \BibitemOpen
  \bibfield  {author} {\bibinfo {author} {\bibfnamefont {F.}~\bibnamefont
  {Garofalo}}, \bibinfo {author} {\bibfnamefont {O.}~\bibnamefont {Richmond}},
  \bibinfo {author} {\bibfnamefont {W.~F.}\ \bibnamefont {Domis}},\ and\
  \bibinfo {author} {\bibfnamefont {F.}~\bibnamefont {von Gemmingen}},\
  }\bibfield  {title} {\bibinfo {title} {{Strain-Time, Rate-Stress and
  Rate-Temperature Relations during Large Deformations in Creep}},\ }\href
  {https://doi.org/10.1243/PIME_CONF_1963_178_010_02} {\bibfield  {journal}
  {\bibinfo  {journal} {Proceedings of the Institution of Mechanical Engineers,
  Conference Proceedings}\ }\textbf {\bibinfo {volume} {178}},\ \bibinfo
  {pages} {31} (\bibinfo {year} {1963})}\BibitemShut {NoStop}%
\bibitem [{\citenamefont {Stang}(2018)}]{Stang2018}%
  \BibitemOpen
  \bibfield  {author} {\bibinfo {author} {\bibfnamefont {E.~T.}\ \bibnamefont
  {Stang}},\ }\emph {\bibinfo {title} {{Constitutive Modeling of Creep in
  Leaded and Lead-Free Solder Alloys Using Constant Strain-Rate Tensile
  Testing}}},\ \href@noop {} {\bibinfo {type} {Master of science in mechanical
  engineering}},\ \bibinfo  {school} {Wright State University} (\bibinfo {year}
  {2018})\BibitemShut {NoStop}%
\bibitem [{\citenamefont {Shin}\ \emph {et~al.}(2015)\citenamefont {Shin},
  \citenamefont {Park}, \citenamefont {Han}, \citenamefont {Sastry},\ and\
  \citenamefont {Lu}}]{Shin2015}%
  \BibitemOpen
  \bibfield  {author} {\bibinfo {author} {\bibfnamefont {H.}~\bibnamefont
  {Shin}}, \bibinfo {author} {\bibfnamefont {J.}~\bibnamefont {Park}}, \bibinfo
  {author} {\bibfnamefont {S.}~\bibnamefont {Han}}, \bibinfo {author}
  {\bibfnamefont {A.~M.}\ \bibnamefont {Sastry}},\ and\ \bibinfo {author}
  {\bibfnamefont {W.}~\bibnamefont {Lu}},\ }\bibfield  {title} {\bibinfo
  {title} {{Component-/structure-dependent elasticity of solid electrolyte
  interphase layer in Li-ion batteries: Experimental and computational
  studies}},\ }\href {https://doi.org/10.1016/j.jpowsour.2014.11.120}
  {\bibfield  {journal} {\bibinfo  {journal} {Journal of Power Sources}\
  }\textbf {\bibinfo {volume} {277}},\ \bibinfo {pages} {169} (\bibinfo {year}
  {2015})}\BibitemShut {NoStop}%
\bibitem [{\citenamefont {Chai}\ \emph {et~al.}(2021)\citenamefont {Chai},
  \citenamefont {Jia}, \citenamefont {Hu}, \citenamefont {Jin}, \citenamefont
  {Jin}, \citenamefont {Ju}, \citenamefont {Yan}, \citenamefont {Ji},\ and\
  \citenamefont {Wan}}]{Chai2021}%
  \BibitemOpen
  \bibfield  {author} {\bibinfo {author} {\bibfnamefont {Y.}~\bibnamefont
  {Chai}}, \bibinfo {author} {\bibfnamefont {W.}~\bibnamefont {Jia}}, \bibinfo
  {author} {\bibfnamefont {Z.}~\bibnamefont {Hu}}, \bibinfo {author}
  {\bibfnamefont {S.}~\bibnamefont {Jin}}, \bibinfo {author} {\bibfnamefont
  {H.}~\bibnamefont {Jin}}, \bibinfo {author} {\bibfnamefont {H.}~\bibnamefont
  {Ju}}, \bibinfo {author} {\bibfnamefont {X.}~\bibnamefont {Yan}}, \bibinfo
  {author} {\bibfnamefont {H.}~\bibnamefont {Ji}},\ and\ \bibinfo {author}
  {\bibfnamefont {L.-J.}\ \bibnamefont {Wan}},\ }\bibfield  {title} {\bibinfo
  {title} {{Monitoring the mechanical properties of the solid electrolyte
  interphase (SEI) using electrochemical quartz crystal microbalance with
  dissipation}},\ }\href {https://doi.org/10.1016/j.cclet.2020.09.008}
  {\bibfield  {journal} {\bibinfo  {journal} {Chinese Chemical Letters}\
  }\textbf {\bibinfo {volume} {32}},\ \bibinfo {pages} {1139} (\bibinfo {year}
  {2021})}\BibitemShut {NoStop}%
\bibitem [{\citenamefont {Edgeworth}\ \emph {et~al.}(1984)\citenamefont
  {Edgeworth}, \citenamefont {Dalton},\ and\ \citenamefont
  {Parnell}}]{Edgeworth1984}%
  \BibitemOpen
  \bibfield  {author} {\bibinfo {author} {\bibfnamefont {R.}~\bibnamefont
  {Edgeworth}}, \bibinfo {author} {\bibfnamefont {B.~J.}\ \bibnamefont
  {Dalton}},\ and\ \bibinfo {author} {\bibfnamefont {T.}~\bibnamefont
  {Parnell}},\ }\bibfield  {title} {\bibinfo {title} {{The pitch drop
  experiment}},\ }\href {https://doi.org/10.1088/0143-0807/5/4/003} {\bibfield
  {journal} {\bibinfo  {journal} {European Journal of Physics}\ }\textbf
  {\bibinfo {volume} {5}},\ \bibinfo {pages} {198} (\bibinfo {year}
  {1984})}\BibitemShut {NoStop}%
\bibitem [{\citenamefont {Sutardja}\ and\ \citenamefont
  {Oldham}(1989)}]{Sutardja1989}%
  \BibitemOpen
  \bibfield  {author} {\bibinfo {author} {\bibfnamefont {P.}~\bibnamefont
  {Sutardja}}\ and\ \bibinfo {author} {\bibfnamefont {W.}~\bibnamefont
  {Oldham}},\ }\bibfield  {title} {\bibinfo {title} {{Modeling of stress
  effects in silicon oxidation}},\ }\href {https://doi.org/10.1109/16.43661}
  {\bibfield  {journal} {\bibinfo  {journal} {IEEE Transactions on Electron
  Devices}\ }\textbf {\bibinfo {volume} {36}},\ \bibinfo {pages} {2415}
  (\bibinfo {year} {1989})}\BibitemShut {NoStop}%
\bibitem [{\citenamefont {Senez}\ \emph {et~al.}(1994)\citenamefont {Senez},
  \citenamefont {Collard}, \citenamefont {Baccus}, \citenamefont {Brault},\
  and\ \citenamefont {Lebailly}}]{Senez1994}%
  \BibitemOpen
  \bibfield  {author} {\bibinfo {author} {\bibfnamefont {V.}~\bibnamefont
  {Senez}}, \bibinfo {author} {\bibfnamefont {D.}~\bibnamefont {Collard}},
  \bibinfo {author} {\bibfnamefont {B.}~\bibnamefont {Baccus}}, \bibinfo
  {author} {\bibfnamefont {M.}~\bibnamefont {Brault}},\ and\ \bibinfo {author}
  {\bibfnamefont {J.}~\bibnamefont {Lebailly}},\ }\bibfield  {title} {\bibinfo
  {title} {{Analysis and application of a viscoelastic model for silicon
  oxidation}},\ }\href {https://doi.org/10.1063/1.357450} {\bibfield  {journal}
  {\bibinfo  {journal} {Journal of Applied Physics}\ }\textbf {\bibinfo
  {volume} {76}},\ \bibinfo {pages} {3285} (\bibinfo {year}
  {1994})}\BibitemShut {NoStop}%
\bibitem [{\citenamefont {Ojovan}(2008)}]{Ojovan2008}%
  \BibitemOpen
  \bibfield  {author} {\bibinfo {author} {\bibfnamefont {M.~I.}\ \bibnamefont
  {Ojovan}},\ }\bibfield  {title} {\bibinfo {title} {{Viscosity and Glass
  Transition in Amorphous Oxides}},\ }\href
  {https://doi.org/10.1155/2008/817829} {\bibfield  {journal} {\bibinfo
  {journal} {Advances in Condensed Matter Physics}\ }\textbf {\bibinfo {volume}
  {2008}},\ \bibinfo {pages} {1} (\bibinfo {year} {2008})}\BibitemShut
  {NoStop}%
\bibitem [{\citenamefont {Schwan}\ \emph {et~al.}(2020)\citenamefont {Schwan},
  \citenamefont {Nava},\ and\ \citenamefont {Mangolini}}]{Schwan2020}%
  \BibitemOpen
  \bibfield  {author} {\bibinfo {author} {\bibfnamefont {J.}~\bibnamefont
  {Schwan}}, \bibinfo {author} {\bibfnamefont {G.}~\bibnamefont {Nava}},\ and\
  \bibinfo {author} {\bibfnamefont {L.}~\bibnamefont {Mangolini}},\ }\bibfield
  {title} {\bibinfo {title} {{Critical barriers to the large scale
  commercialization of silicon-containing batteries}},\ }\href
  {https://doi.org/10.1039/D0NA00589D} {\bibfield  {journal} {\bibinfo
  {journal} {Nanoscale Advances}\ }\textbf {\bibinfo {volume} {2}},\ \bibinfo
  {pages} {4368} (\bibinfo {year} {2020})}\BibitemShut {NoStop}%
\bibitem [{\citenamefont {Durdel}\ \emph {et~al.}(2023)\citenamefont {Durdel},
  \citenamefont {Friedrich}, \citenamefont {H{\"{u}}sken},\ and\ \citenamefont
  {Jossen}}]{Durdel2023}%
  \BibitemOpen
  \bibfield  {author} {\bibinfo {author} {\bibfnamefont {A.}~\bibnamefont
  {Durdel}}, \bibinfo {author} {\bibfnamefont {S.}~\bibnamefont {Friedrich}},
  \bibinfo {author} {\bibfnamefont {L.}~\bibnamefont {H{\"{u}}sken}},\ and\
  \bibinfo {author} {\bibfnamefont {A.}~\bibnamefont {Jossen}},\ }\bibfield
  {title} {\bibinfo {title} {{Modeling Silicon-Dominant Anodes:
  Parametrization, Discussion, and Validation of a Newman-Type Model}},\ }\href
  {https://doi.org/10.3390/batteries9110558} {\bibfield  {journal} {\bibinfo
  {journal} {Batteries}\ }\textbf {\bibinfo {volume} {9}},\ \bibinfo {pages}
  {558} (\bibinfo {year} {2023})}\BibitemShut {NoStop}%
\bibitem [{\citenamefont {Plett}(2004)}]{Plett2004}%
  \BibitemOpen
  \bibfield  {author} {\bibinfo {author} {\bibfnamefont {G.~L.}\ \bibnamefont
  {Plett}},\ }\bibfield  {title} {\bibinfo {title} {{Extended Kalman filtering
  for battery management systems of LiPB-based HEV battery packs}},\ }\href
  {https://doi.org/10.1016/j.jpowsour.2004.02.032} {\bibfield  {journal}
  {\bibinfo  {journal} {Journal of Power Sources}\ }\textbf {\bibinfo {volume}
  {134}},\ \bibinfo {pages} {262} (\bibinfo {year} {2004})}\BibitemShut
  {NoStop}%
\bibitem [{\citenamefont {Graells}\ \emph {et~al.}(2020)\citenamefont
  {Graells}, \citenamefont {Trimboli},\ and\ \citenamefont
  {Plett}}]{Graells2020}%
  \BibitemOpen
  \bibfield  {author} {\bibinfo {author} {\bibfnamefont {C.~P.}\ \bibnamefont
  {Graells}}, \bibinfo {author} {\bibfnamefont {M.~S.}\ \bibnamefont
  {Trimboli}},\ and\ \bibinfo {author} {\bibfnamefont {G.~L.}\ \bibnamefont
  {Plett}},\ }\bibfield  {title} {\bibinfo {title} {{Differential hysteresis
  models for a silicon-anode Li-ion battery cell}},\ }in\ \href
  {https://doi.org/10.1109/ITEC48692.2020.9161591} {\emph {\bibinfo {booktitle}
  {2020 IEEE Transportation Electrification Conference \& Expo (ITEC)}}},\
  Vol.~\bibinfo {volume} {1}\ (\bibinfo  {publisher} {IEEE},\ \bibinfo {year}
  {2020})\ pp.\ \bibinfo {pages} {175--180}\BibitemShut {NoStop}%
\bibitem [{\citenamefont {Wycisk}\ \emph {et~al.}(2022)\citenamefont {Wycisk},
  \citenamefont {Oldenburger}, \citenamefont {Stoye}, \citenamefont
  {Mrkonjic},\ and\ \citenamefont {Latz}}]{Wycisk2022}%
  \BibitemOpen
  \bibfield  {author} {\bibinfo {author} {\bibfnamefont {D.}~\bibnamefont
  {Wycisk}}, \bibinfo {author} {\bibfnamefont {M.}~\bibnamefont {Oldenburger}},
  \bibinfo {author} {\bibfnamefont {M.~G.}\ \bibnamefont {Stoye}}, \bibinfo
  {author} {\bibfnamefont {T.}~\bibnamefont {Mrkonjic}},\ and\ \bibinfo
  {author} {\bibfnamefont {A.}~\bibnamefont {Latz}},\ }\bibfield  {title}
  {\bibinfo {title} {{Modified Plett-model for modeling voltage hysteresis in
  lithium-ion cells}},\ }\href {https://doi.org/10.1016/j.est.2022.105016}
  {\bibfield  {journal} {\bibinfo  {journal} {Journal of Energy Storage}\
  }\textbf {\bibinfo {volume} {52}},\ \bibinfo {pages} {105016} (\bibinfo
  {year} {2022})}\BibitemShut {NoStop}%
\end{thebibliography}%

\end{document}